\documentclass[12pt,a4paper]{article}
\usepackage{graphicx,amssymb}
\usepackage{graphicx, array, epsfig, subfigure}
\usepackage{multirow}
\usepackage{subfigure}  % use for side-by-side figures
\newcommand{\be}{\begin{equation}}
\newcommand{\ee}{\end{equation}}
\newcommand{\bea}{\begin{eqnarray}}
\newcommand{\eea}{\end{eqnarray}}

\catcode`@=12

\begin{document}
\begin{center}
{\bf IceCube PeV Neutrino Events from the Decay of Superheavy Dark Matter; 
an Analysis}\\
%{\bf Mass and Life Time of Heavy Dark Matter Decaying into IceCube PeV 
%Neutrinos}\\
\vspace{1cm}
{{\bf Madhurima Pandey$^{a}$} \footnote {email: madhurima.pandey@saha.ac.in},
{\bf Debasish Majumdar$^{a}$} \footnote {email: debasish.majumdar@saha.ac.in}},\\
{\normalsize \it $^a$Astroparticle Physics and Cosmology Division,}  \\
{\normalsize \it Saha Institute of Nuclear Physics, HBNI}  \\
{\normalsize \it 1/AF Bidhannagar, Kolkata 700064, India } \\
\vspace{0.25cm}
{\bf Ashadul Halder$^{b}$} \footnote{email: ashadul.halder@gmail.com}\\
{\normalsize \it $^{b}$Department of Physics, St. Xavier's College,} \\
{\normalsize \it 30, Mother Teresa Sarani, Kolkata - 700016, India}  \\
\vspace{1cm}
\end{center}

\begin{center}
{\bf Abstract}
\end{center}

{\small  
Considering the ultrahigh energy (UHE) neutrino events reported by IceCube in the PeV regime to have originated from the decay of superheavy dark matter, 
the IceCube UHE neutrino events are analysed and the best fit values of the two parameters namely the mass of the superheavy dark matter and its decay lifetime are obtained. The theoretical astrophysical flux is also included in the 
analysis. We find that while the neutrino events in the energy range 
$\sim$ 60 TeV-$\sim$ 120 TeV appears to have astrophysical origin, 
the events in the 
energy range $\sim 1.2 \times 10^5$ GeV - $\sim 5 \times 10^7$ GeV can be well 
described 
from the superheavy dark matter decay hypothesis. We also find that although 
hadronic decay channel of the superheavy dark matter can well explain the 
events in the energy range $\sim 1.2 \times 10^5$ GeV - $\sim 5 \times 10^6$ 
GeV, the
higher energy regime higher than this range can be addressed only when 
the leptonic decay channel is considered.}
\newpage
\section{Introduction}
The Icecube (IC) detector at the south pole which uses the southpole ice as 
the detector material for the neutrinos has so far reported several neutrino 
events with energies ranging between few hundreds of GeV upto the 
ultrahigh energy regime of tens of PeV. The events in the TeV-PeV range 
enables one to probe the neutrinos from extragalactic sources that could 
produce such high energy neutrinos. This also helps to identify and to 
ascertain the nature and properties of such possible high energy sources. 
The event data at IceCube for the neutrino energies greater that 20 TeV 
are categorized as high energy starting event (HESE) data. The possibility 
of sources for such data can be of wide range. e.g. 
Supernova Remnants (SNR) \cite{snr}, Active Galactic Nuclei (AGN) 
\cite{agn1,agn2,agn3}, Gamma Ray Bursts (GRBs) \cite{grb} etc. 
There are also other suggestions in the literature that these UHE neutrinos 
(in and around PeV region) would have originated from the decay of 
very heavy dark matter \cite{anisimov}-\cite{icecube25}. 

In this work we  
consider the latter possibility mentioned above that the decay of 
superheavy dark matter (SHDM) 
could produce the neutrinos detected by the IceCube detector in the energy 
range $\sim$ 60 Tev-$\sim$ 50 PeV. 
These SHDMs can produce neutrinos by rare long lived decay processes. They are 
most likely non-thermal in nature as they could have been created in the early 
Universe by spontaneous symmetry breaking or through the process of 
gravitational creation \cite{univ1,paolo}. The IC Collaboration fitted their 
data of 82 events (HESE data \cite{icrc}) that include both shower and track 
events to obtaine a power law spectrum for the detected neutrino flux. To 
this end, they fit first an unbroken power law spectrum of the form 
$\sim E^{- \gamma}$ where the fitted value of $\gamma$ is found to be 
$2.92_{-0.29}^{+0.33}$. The HESE data is referred to the one where the 
IC Collaboration obtained a flux $\sim E^{-2.9}$ after a single power law 
fit. But a fit of the data between the energy range $\sim$ 120 TeV to $\sim$ 
5 PeV yielded a power law different from the HESE fit ($\gamma \sim 2.9$) 
indicating the possibility of a second component. This component had been 
represented as a pink band (specifying 1$\sigma$ uncertainty) in Fig. 2 
of Ref. \cite{icrc} where IC collaboration furnished their event data for 
the region(s) now in discussion. Beyond the 
energy $\sim$ 5 PeV the data points as shown in Fig. 2 of \cite{icrc} are 
found to have no lower bound and these data do not appear to follow the 
fitted spectrum. In this work we consider however the event data points given 
in Fig. 2 of \cite{icrc} from $\sim$ 60 TeV upto $\sim$ 50 PeV.

In an earlier work \cite{mpplb}, the energy regime only within the pink band has
been considered 
and it has been shown that this part can be well explained by considering a
superheavy dark matter decaying to neutrinos only through hadronic channel. 
In the present work where we have made a $chi^2$ analyses for the event data
within a larger energy regime of $\sim 60$ TeV - $\sim$ 50 PeV, we show that
in order to explain the apparent 
nature of the flux beyond $\sim$ 5 PeV one needs to include superheavy 
dark matter 
decay through leptonic channel. We also demonstrate that the 
two event data 
points in the energy range $\sim$ 60 TeV - $\sim$ 120 TeV in Fig 2 of 
\cite{icrc}, can be 
best explained if these two are considered to have astrophysical origin. 
Our analysis of the whole range of events (from $\sim$ 60 TeV to $\sim$ 50 
PeV) suggests that this range seems to have three parts in terms 
of the possible origins of neutrino events. 
The energy range spanning between 
$\sim$ 60 TeV - $\sim$ 120 TeV containing two event data points is the 
astrophysical 
component and both the second and third components could have originated 
from the decay of SHDM. Among the latter two components the event data ranging 
between $\sim 1.2 \times 10^{5}$ GeV - $\sim 5 \times 10^{6}$ GeV are from the 
hadronic cascade decay of SHDM while the event 
range $\sim 5 \times 10^{6}$ GeV - $\sim 5 \times 10^{7}$ GeV  are from the 
leptonic decay channel of the same SHDM.

%We also take into account the 3-flavour neutrino oscillation/suppression of 
%the neutrinos on reaching the Earth in our analysis. This may also be 
%mentioned here that we have repeated our analysis considering the 
%4 (3 active and one sterile) flavour scenario for  
%the oscillation/suppression part, where the oscillation is assumed 
%to have realised between four flavours. 
%But we find no significant change in results with those obtained 
%for usual 3-flavour case.

In the present analysis, besides incorporating the possible astrophysical
origin of neutrinos (diffuse flux), both the 
hadronic and leptonic channels of the decay cascade of these SHDMs 
that finally produce the UHE neutrinos are included along with the 
oscillations/suppressions that the neutrinos of a certain flavour would suffer
while traversing the 
astronomical distance to reach the Earth. We consider three active 
flavour neutrinos as also the four (3+1) neutrino scheme, where a 
sterile neutrino is added to the usual three flavour scenario and 
compare our results.  
But we find no significant change in results with those obtained 
for usual 3-flavour case.

The study of the IceCube data \cite{icrc} in this 
regime shows that 
there are two points at energies $\sim$ 60 TeV - $\sim$ 120 TeV and 
eight number of points 
are in the energy regime greater than $\sim$ 120 TeV. Among the latter data 
set three data points do not have any experimental lower limit. The 
first two data 
are supposed to be of astrophysical origin \cite{upl}. 
From the analysis presented in this work, it appears that the 
SHDM decay considerations 
for UHE neitrino production is most relevant for the neutrinos 
with energies $ >$ 120 TeV (\cite{icrc}). 

The whole set is then fitted with 
the corresponding data points given by IceCube Collaboration to obtain 
From the $\chi^2$ analysis presented in this work,
the best fit values of the mass of the SHDM and its decay lifetime are
also obtained. 
%3-flavour and the 4-flavour 
%cases and the best fit values for both the cases are obtained. 

The paper is organised as follows. In Section. 2 we briefly describe the 
superheavy dark matter decay process for both hadronic and leptonic decay 
channels. UHE neutrino flux in Section. 3 has two subsections. 
In Subsection. 3.1 we consider the neutrino fluxes 
from the astrophysical sources while Subsection 3.2 contains the UHE 
neutrino fluxes originated from the superheavy dark matter decay. Section. 
4 deals with the modified UHE neutrino fluxes at the Earth for both 
3-flavour and 4-flavour framework. The calculational results are discussed 
in Section. 5. Finally, we summarise the paper in Section. 6.

\section{Superheavy Dark Matter Decays}
The decay of superheavy dark matter particles (SHDMs) that 
could be produced in the early Universe proceed via the cascading of 
QCD partons.
For the case of the decay of SHDM particles 
with masses $m_{\chi}$ much larger than the electroweak scale 
$m_{\chi} \gg m_W$, the electroweak cascade occurs in addition to the 
QCD cascade \cite{bere1,bere2}. The production mechanism of the hadronic 
QCD spectrum including supersymmetric QCD cascade depends on two 
different methods. One of these two most effective methods is the 
Monte Carlo (MC) simulation \cite{bere3,bere4} while the other one is the 
Dokshitzer-Gribov-Lipatov-Altarelli-Parisi (DGLAP) equation 
\cite{bere4}-\cite{cyrille2}, which describes the evolution of the 
fragmentation function. We obtain the neutrino spectrum/flux as the final 
product of the numerical evolution of the DGLAP equations and the MC studies 
and this spectrum/flux is used in this work to explain the IceCube events 
in ultrahigh energy 
regime. The whole process of the production of neutrino spectrum can be 
described by the two decay channels namely, 
the hadronic and the leptonic decay channels.

\subsection{Hadronic Decay Channels of SHDM}
QCD cascade from the decay of superheavy particles plays a significant 
role to describe the production of hadrons. It is asserted that 
even though the QCD coupling is small, the cascading in QCD parton appears 
due to the enhancement of the parton splitting in the presence of large 
logaritms for soft parton emission. The electroweak radiative corrections 
can also be dominated by similar logarithms. In the case of QCD cascade, 
we use the numerical code \cite{bere4} for the evolution of the DGLAP 
equations. 
A similar kind of approach can be adopted by the electroweak radiative 
corrections at the TeV energy scale or above \cite{ew}-\cite{ew5} 
valid for spontaneously broken gauge group. In this section, 
we discuss in particular the hadronic decay channel 
$\chi \rightarrow q \bar{q}$, where $q$ indicates a quark with a 
flavour. In this decay channel, after the perturbative evolution of the 
QCD cascade, the partons are hadronized and finally, as the end product, 
the leptons are obtained by the subsequent decay of the unstable hadrons. 
In comparison to the other theoretical uncertainties the effect of 
electroweak radiative corrections on the cascade development are insignificant.

The neutrino spectrum can be written as \cite{kuz}
\bea
\displaystyle\frac {dN_{\nu}} {dx} &=& 2R \int_{xR}^{1} \displaystyle \frac
{dy} {y} D^{{\pi}^{\pm}} (y) + 2 \int_{x}^{1} \displaystyle \frac {dz} {z}
f_{\nu_i}
\left (\displaystyle\frac {y} {z} \right) D^{{\pi}^{\pm}}_i (z)\,\, ,
\label{form1}
\eea
where $D^{\pi}_i (x,s)$ ($\equiv [D_{q}^{\pi} (x,s) + D_{g}^{\pi} (x,s)]$ 
is a fragmentation function of the pions from a parton $i (=q(=u,d,s, ...),g)$. 
The total decay spectrum $F^h (x,s)$ can be obtained by the 
summation of the contributions of all possible parton (quarks, antiquarks 
and gluons) fragmentation functions ($D^{\pi}_i (x,s)$), where 
$x (\equiv 2E/m_\chi)$, a dimensionless quantity, defines the fraction 
of energy transferred to the hadron 
and $\sqrt{s}$ is the centre of mass energy. 
In our calculation we consider only the contribution of pion decays and 
the contribution ($\sim$ 10\%) from other mesons are neglected following 
Ref. \cite{bere4}. In Eq. (\ref{form1}), $R = \displaystyle\frac {1} {1-r}$, 
where
$r = (m_\mu/m_\pi)^2 \simeq 0.573$ and the functions $f_{\nu_i} (x)$ are 
given as \cite{kelner}
\bea
f_{\nu_i} (x) &=& g_{\nu_i} (x) \Theta (x-r) +(h_{\nu_i}^{(1)} (x) +
h_{\nu_i}^{(2)} (x))\Theta(r-x) \,\, , \nonumber\\
g_{\nu_\mu} (x) &=& \displaystyle\frac {3-2r} {9(1-r)^2} (9x^2 - 6\ln{x} -4x^3 -5)\,\, , \nonumber\\
h_{\nu_\mu}^{(1)} (x) &=& \displaystyle\frac {3-2r} {9(1-r)^2} (9r^2 - 6\ln{r} -
4r^3 -5)\,\, , \nonumber\\
h_{\nu_\mu}^{(2)} (x) &=& \displaystyle\frac {(1+2r)(r-x)} {9r^2} [9(r+x) -
4(r^2+rx+x^2)]\,\, , \nonumber\\
g_{\nu_e} (x) &=& \displaystyle\frac {2} {3(1-r)^2} [(1-x) (6(1-x)^2 +
r(5 + 5x - 4x^2)) + 6r\ln{x}]\,\, \nonumber\\
h_{\nu_e}^{(1)} (x) &=& \displaystyle\frac {2} {3(1-r)^2} [(1-r)
(6-7r+11r^2-4r^3) + 6r \ln{r}]\,\, , \nonumber\\
h_{\nu_e}^{(2)} (x) &=& \displaystyle\frac {2(r-x)} {3r^2} (7r^2 - 4r^3 +7xr
-4xr^2 - 2x^2 - 4x^2r)\,\, .
\label{form2}
\eea
\subsection{Leptonic Decay Channels of SHDM}
The development of electroweak cascade can be illustrated by considering a 
tree level decay of superheavy particle $\chi$ with mass 
$m_{\chi} \leq m_{\rm GUT}$ to leptons. According to the $Z$-burst model, 
$\chi$ particles are decaying into $\bar{l} l$ pairs and 
$\chi \rightarrow \bar{\nu} \nu$ is the corresponding decay mode. For 
$m_{\chi} \gg m_z$ ($m_z$ being the $z$ boson mass), considering the 
available momentum flow, 
($Q^2 \leq \frac {m_{\chi}^2} {4}$) we can neglect the mass of the $Z$ boson. 
The smallness of the QCD coupling can be compensated by a very large logarithms $ln^{2} (m_{\chi}^2/m_z^2)$, which is generated for soft or collinear 
singularities. Similarly for $m_{\chi} \gg m_W$ ($m_W$ being the mass of 
$W$ boson), due to  
large logarithms the perturbation theory is no more valid and this initiates 
developing of the electroweak cascade, very similar to that for known 
QCD cascade. There can be a
mutual transmutation of electroweak and QCD cascades because 
the electroweak gauge bosons also split into quarks. This will modify the 
hadronic spectra to a limited extent while on the other hand the splitting 
like $W \rightarrow \bar{\nu} \nu$ contributes to the electroweak part of the 
cascade. In order to explain the effects of the SHDM particles 
decaying into the neutrinos as the final product via the leptonic decay 
channels, the MC simulations for both the QCD part \cite{bere3} and the 
electroweak cascade \cite{bere6} have been performed.  

\section{Ultrahigh Energy Neutrino Flux}
In this work, we consider the neutrino flux with energies above $\sim$ 60 TeV. 
The present analysis has been performed for the high energetic IceCube 
events by considering 
two different components of the neutrino flux, namely the 
astrophysical neutrino flux and the neutrino flux from the SHDM decay.

\subsection{Astrophysical Neutrino Flux}
Numerous astrophysical sources can produce high energy neutrinos through 
their highly energetic particle acceleration mechanism of protons, 
where the latter interact with 
themselves ($pp$ interactions) or with photons ($p \gamma$ interactions) 
to finally produce neutrinos. 
Distant ultrahigh energy (UHE) sources like extragalctic Supernova Remnants 
(SNR) \cite{snr}, Active Galactic Nuclei (AGN) \cite{agn1,agn2,agn3}, 
Gamma Ray Bursts (GRBs) \cite{grb}
etc. are proposed as the source of IceCube neutrino induced muon events in 
UHE regime. In the particle acceleration process, a high energetic shock 
wave generates and progresses outwards with energies as high as $\sim 10^{53}$ 
ergs in the form of fireball. The interactions between the protons and the 
photons inside such a fireball produce pions, while these pions decay to 
finally produce UHE 
neutrinos. In this work, we consider the contribution of the astrophysical 
neutrino flux as the source neutrino flux in the $\sim$ 60 - $\sim$ 120 TeV 
energy range.

In order to consider the acceleration mechanism related to the 
astrophysical sources, the isotropic fluxes for the neutrinos are 
estimated by an Unbroken Power Law (UPL) after summing over 
all the possible sources and is given as \cite{upl}
\bea
E_{\nu}^2 \displaystyle\frac {d \phi'_{\nu_{\rm Ast}}} {d E_\nu} (E_\nu) &=& 
N \left (\displaystyle \frac {E_\nu} {100 {\rm TeV}} \right)^{-\gamma}\,\, 
{\rm GeV}\,\, {\rm cm^{-2}}\,\, {\rm s^{-1}}\,\, {\rm sr^{-1}}\,\, ,
\label{flux}
\eea
where $N$ represents the normalization factor of the flux and $\gamma$ is 
the spectral index. We have chosen the values of $N$ and $\gamma$ for our 
analyses as $1 \times 10^{-8}$ and $1.0$ respectively for UPL \cite{upl}. 
Assuming the neutrinos are produced in the flavour ratio 1:2:0, the flux for 
$\nu_e$ in this case at source will be 
$\displaystyle\frac {1} {3} \displaystyle\frac {d \phi'_{\nu_{\rm Ast}}} {d E_\nu} = \displaystyle\frac {d \phi_{\nu_{\rm Ast}}} {d E_\nu}$.

Here we mention that for the astrophysical flux we also adopt the 
power law spectrum given by IceCube Collaboration in Ref. \cite{icrc} 
(Fig. 2 of \cite{icrc}). This is of the form 
\bea 
E_{\nu}^2 \frac{d \phi} {dE_\nu} (E_\nu) &=& 2.46 \pm 0.8 \times 10^{-8} 
(E/100 {\rm TeV})^{-0.92}\,\, {\rm GeV}\,\, {\rm cm}^{-2}\,\, {\rm s}^{-1}\,\, 
{\rm sr}^{-1}\,\,. 
\label{icflux}
\eea
Thus our analysis is done for each of the both astrophysical fluxes given here.

\subsection{Neutrino Flux from Superheavy Dark Matter Decay}
The neutrino flux from superheavy dark matter decay has two components namely a 
galactic component and the other an extragalactic component. The galactic 
neutrino flux from the decay of superheavy dark matter with 
mass $m_{\chi}$ and decay lifetime $\tau$ can be written as
\bea
\displaystyle\frac {d\Phi_{\rm G}} {dE_\nu} (E_\nu) &=& \displaystyle\frac {1}
{4\pi m_\chi \tau} \int_{V} \displaystyle\frac {\rho_\chi (R[r])}
{4\pi r^2} \displaystyle\frac {dN} {dE} (E,l,b) dV \,\, ,
\label{form3}
\eea
where the neutrino spectrum from decaying superheavy dark matter 
particle is defined as $\displaystyle\frac {dN} {dE} (E,l,b)$, $l$ and $b$ 
are the galactic coordinates. In the above,  
$\rho_\chi (R[r])$ is the dark matter density, which is a function of the 
distance ($R$) from the Galactic Center and $r$ indicates the distance from 
the Earth. We adopt the Navarro-Frenk-White (NFW) profile for the 
dark matter density \cite{nfw1,nfw2} in this work. The integration is 
made over the 
Milky Way halo for which the maximum value of $R$ is chosen as 
$R_{\rm max} = 260$ Kpc \cite{milky}.   

The isotropic extragalctic neutrino flux from similar decay is given as  
\bea
\displaystyle\frac {d\Phi_{\rm EG}} {dE_\nu} (E_\nu) &=& \displaystyle\frac {1}
{4\pi m_\chi \tau} \int_{0}^{\infty} \displaystyle\frac {\rho_0 c /H_0} 
{\sqrt{\Omega_m (1+z^3) + (1-\Omega_m)}} \displaystyle\frac {dN} {dE} [E(1+z)]
dz \,\, .
\label{form4}
\eea
In the above equation (Eq. (\ref{form4})), 
$c/H_0 = 1.37 \times 10^{28}\,\, {\rm cm}$ signifies the Hubble radius and
$\rho_0\,\, (= 1.15 \times 10^{-6}$ GeV/cm$^3$) is the average 
cosmological dark matter density at the redshift $z = 0$ (present epoch). 
The contribution of the matter density to the energy density of the 
Universe in units of the critical energy density is defined as 
$\Omega_m = 0.316$. The injected neutrino energy spectrum obtained from the 
decay of superheavy particles is denoted as $\frac {dN} {dE_\nu}$, which 
is a function of the particle energy shift $z$, $E(z) = (1+z)E$. For both 
the galactic and extragalactic neutrino fluxes it is assumed that they 
reach Earth in the ratio 1:1:1 for three neutrino flavours 
($\nu_e,\nu_\mu,\nu_\tau$). Therefore at source the 
$\nu_e$ flux can be taken to be Eq. (\ref{form3}) and Eq. (\ref{form4}) 
for galactic and extragalactic cases respectively. This is also to note that 
each of the fluxes $\displaystyle\frac {d\Phi_{\rm EG}} {dE} (E_\nu)$ and 
$\displaystyle\frac {d\Phi_{\rm G}} {dE} (E_\nu)$ in fact has two 
components namely the one that is of hadronic origin and the other which 
is obtained from leptonic decay channel. Therefore 
$\displaystyle\frac {d\Phi_{\rm EG}} {dE} (E_\nu) = 
\left (\displaystyle\frac {d\Phi_{\rm EG}} {dE} (E_\nu) \right)_{\rm had} + 
\left (\displaystyle\frac {d\Phi_{\rm EG}} {dE} (E_\nu) \right)_{\rm lep}$ and 
$\displaystyle\frac {d\Phi_{\rm G}} {dE} (E_\nu) = 
\left (\displaystyle\frac {d\Phi_{\rm G}} {dE} (E_\nu) \right)_{\rm had} + 
\left (\displaystyle\frac {d\Phi_{\rm G}} {dE} (E_\nu) \right)_{\rm lep}$. 

Thus the total electron neutrino flux at the source, (diffuse astrophysical
sources and the decay of superheavy dark matter), can be written as
\bea
\phi^{\rm th} (E_\nu) &=&  \displaystyle\frac {d \phi_{\nu_{\rm Ast}}} 
{d E_\nu} (E_\nu) +
\left (\displaystyle\frac {d\Phi_{\rm EG}} {dE} (E_\nu) \right)_{\rm had} +
\left (\displaystyle\frac {d\Phi_{\rm G}} {dE} (E_\nu) \right)_{\rm had} + 
\left (\displaystyle\frac {d\Phi_{\rm EG}} {dE} (E_\nu) \right)_{\rm lep} + 
\nonumber \\
&&
\left (\displaystyle\frac {d\Phi_{\rm G}} {dE} (E_\nu) \right)_{\rm lep} \,\, ,
\label{form5}
\eea
where for the first term on the R.H.S., two analytical forms are adopted
as discussed in Sect. 3.1. 
\section{Neutrino Oscillations and the Modified Fux}
A neutrino $|\nu_\alpha\rangle$ of flavour $\alpha$ can oscillate to a neutrino  $|\nu_\beta\rangle$ with flavour $\beta$ after traversing a baseline length 
of $L$ and the oscillation probability can be written as \cite{prob}
\bea
P_{\nu_\alpha \rightarrow \nu_\beta} &=& \delta_{\alpha\beta}
- 4\displaystyle\sum_{j>i} U_{\alpha i} U_{\beta i} U_{\alpha j} U_{\beta j}
\sin^2\left (\frac {\pi L} {\lambda_{ij}} \right )\,\, ,
\label{oscprob}
\eea
where $i,j$ indicate the mass indices and $U_{\alpha i}$ etc. denote 
the neutrino mass-flavour mixing matrix elements 
(Pontecorvo-Maki-Nakagawa-Sakata (PMNS) matrix) \cite{pmns}. The flavour 
eigenstate $|\nu_\alpha \rangle$ relates to 
the mass eigenstate $|\nu_i\rangle$ through the relation
\bea
|\nu_\alpha \rangle &=& \displaystyle\sum_{i} U_{\alpha i} |\nu_i \rangle\,\, ,
\label{completeset}
\eea
The oscillation length, which is mentioned in Eq. (\ref{oscprob}), is given 
by
\bea
\lambda_{ij} &=& 2.47\,{\rm Km} \left ( \displaystyle\frac {E} {\rm GeV}
\right ) \left (\displaystyle\frac {{\rm eV}^2} {\Delta m^2_{ij}} \right )\,\,  .
\label{osclen}
\eea
The oscillatory part of the probability equation (Eq. (\ref{oscprob})) is 
averaged to 1/2 due to the long astronomical baseline distance $L$ for the 
present context of UHE neutrinos (${\Delta m^2 L}/{E} \gg 1$), where $E$ is 
the neutrino energy 
and ${\Delta m^2_{ij}}$ being the mass square difference 
 of two neutrinos with mass eigenstates $|\nu_i\rangle$ and $|\nu_j\rangle$. 
Therefore,
\bea
\left \langle \sin^2 \left ( \displaystyle\frac {\pi L}{\lambda_{ij}} \right )
\right \rangle &=& \frac {1}{2}\,\,  .
\label{longbasline}
\eea
With this, the oscillation probability equation (Eq. (\ref{oscprob})) takes 
the form
\bea
P_{\nu_\alpha \rightarrow \nu_\beta} &=& \delta_{\alpha \beta}
- 2 \displaystyle\sum_{j>i} U_{\alpha i} U_{\beta i} U_{\alpha j} U_{\beta j}   \nonumber \\
&=& \delta_{\alpha \beta} - \displaystyle\sum_i U_{\alpha i} U_{\beta i}
\left [\displaystyle\sum_{j\ne i} U_{\alpha j} U_{\beta j} \right ]
\nonumber  \\
&=& \displaystyle\sum_{j} {\mid U_{\alpha j}\mid}^2 {\mid U_{\beta j}\mid}^2\,\, .
\label{oscprob1}
\eea
In the above we use the unitarity condition
\bea
\displaystyle\sum_{i} U_{\alpha i} U_{\beta i} &=& \delta_{\alpha\beta}\,\,  .
\label{unitarity}
\eea
Therefore the flux for each flavour on reaching the Earth can be derived as 
(with the assumption that neutrinos are produced in the ratio 
$\nu_e:\nu_\mu:\nu_\tau = 1:2:0$).
{\small
\bea
\left (\begin{array}{c}
F^3_{\nu_e} \\
F^3_{\nu_\mu} \\
F^3_{\nu_\tau} \end{array} \right)
&=& \left (\begin{array}{ccc}
{\mid {\cal{U}}_{e1} \mid}^2 & {\mid {\cal{U}}_{e2} \mid}^2 &
{\mid {\cal{U}}_{e3} \mid}^2     \\
{\mid {\cal{U}}_{\mu1} \mid}^2 & {\mid {\cal{U}}_{\mu2} \mid}^2 &
{\mid {\cal{U}}_{\mu3} \mid}^2  \\
{\mid {\cal{U}}_{\tau1} \mid}^2 & {\mid {\cal{U}}_{\tau2} \mid}^2 &
{\mid {\cal{U}}_{\tau3} \mid}^2 \end{array} \right)
\left (\begin{array}{ccc}
{\mid {\cal{U}}_{e1} \mid}^2 & {\mid {\cal{U}}_{\mu1} \mid}^2 &
{\mid {\cal{U}}_{\tau1} \mid}^2   \\
{\mid {\cal{U}}_{e2} \mid}^2 & {\mid {\cal{U}}_{\mu2} \mid}^2 &
{\mid {\cal{U}}_{\tau2} \mid}^2   \\
{\mid {\cal{U}}_{e3} \mid}^2 & {\mid {\cal{U}}_{\mu3} \mid}^2 &
{\mid {\cal{U}}_{\tau3} \mid}^2 \end{array} \right)\,\,\nonumber\\
& &\times
\left (\begin{array}{c}
1 \\
2 \\
0 \end{array} \right )
\phi_{\nu_e}\,\, .
\label{cosflux2}
\eea
}
With the unitarity conditions of the PMNS matrix the flux for each flavour 
on reaching the Earth is finally written as
{\small
\bea
F^3_{\nu_e} &=& [ {\mid {\cal{U}}_{e1} \mid}^2 (1 + {\mid {\cal{U}}_{\mu1} \mid}^2 - {\mid {\cal{U}}_{\tau1} \mid}^2 )
+ {\mid {\cal{U}}_{e2} \mid}^2 (1 + {\mid {\cal{U}}_{\mu2} \mid}^2  -
{\mid {\cal{U}}_{\tau2} \mid}^2 ) \nonumber\\
& &  + {\mid {\cal{U}}_{e3} \mid}^2 (1 + {\mid {\cal{U}}_{\mu3} \mid}^2 -
{\mid {\cal{U}}_{\tau3} \mid}^2 )]\phi_{\nu_e}\,\,\, ,\nonumber\\
F^3_{\nu_\mu} &=& [ {\mid {\cal{U}}_{\mu1} \mid}^2 (1 + {\mid {\cal{U}}_{\mu1} \mid}^2 - {\mid {\cal{U}}_{\tau1} \mid}^2 ) + {\mid {\cal{U}}_{\mu2} \mid}^2 (1 + {\mid {\cal{U}}_{\mu2} \mid}^2  -
{\mid {\cal{U}}_{\tau2} \mid}^2 )  \nonumber\\
& &+ {\mid {\cal{U}}_{\mu3} \mid}^2 (1 + {\mid {\cal{U}}_{\mu3} \mid}^2 -
{\mid {\cal{U}}_{\tau3} \mid}^2 )]\phi_{\nu_e}\,\,\, ,\nonumber\\
F^3_{\nu_\tau} &=& [ {\mid {\cal{U}}_{\tau1} \mid}^2 (1 + {\mid {\cal{U}}_{\mu1} \mid}^2 - {\mid {\cal{U}}_{\tau1} \mid}^2 ) + {\mid {\cal{U}}_{\tau2} \mid}^2 (1 + {\mid {\cal{U}}_{\mu2} \mid}^2  - {\mid {\cal{U}}_{\tau2} \mid}^2 )\nonumber\\
& & + {\mid {\cal{U}}_{\tau3} \mid}^2 (1 + {\mid {\cal{U}}_{\mu3} \mid}^2 -
{\mid {\cal{U}}_{\tau3} \mid}^2 )]\phi_{\nu_e}\,\,\, .
\label{3fmuprob}
\eea
}
Where 
{\small
\bea
{\cal{U}} &=& \left ( \begin{array}{ccc}
c_{12}c_{13} & s_{12}c_{13} & s_{13} \\
-s_{12}c_{23}-c_{12}s_{23}s_{13} & c_{12}c_{23}-s_{12}s_{23}s_{13} &
s_{23}c_{13} \\
s_{12}s_{23}-c_{12}c_{23}s_{13} & -c_{12}s_{23}-s_{12}c_{23}s_{13} &
c_{23}c_{13}  \end{array} \right )\,\,  .
\label{pmns3f}
\eea
Proceeding similarly for the 4-flavour oscillation scenario, the flux for 
each flavour on reaching the Earth is obtained as 
{\small
\bea
\left (\begin{array}{c}
F^4_{\nu_e} \\
F^4_{\nu_\mu} \\
F^4_{\nu_\tau} \\
F^4_{\nu_s} \end{array} \right)
&=& \left (\begin{array}{cccc}
{\mid {\tilde{U}}_{e1} \mid}^2 & {\mid {\tilde{U}}_{e2} \mid}^2 &
{\mid {\tilde{U}}_{e3} \mid}^2 & {\mid {\tilde{U}}_{e4} \mid}^2   \\
{\mid {\tilde{U}}_{\mu1} \mid}^2 & {\mid {\tilde{U}}_{\mu2} \mid}^2 &
{\mid {\tilde{U}}_{\mu3} \mid}^2 & {\mid {\tilde{U}}_{\mu4} \mid}^2   \\
{\mid {\tilde{U}}_{\tau1} \mid}^2 & {\mid {\tilde{U}}_{\tau2} \mid}^2 &
{\mid {\tilde{U}}_{\tau3} \mid}^2 & {\mid {\tilde{U}}_{\tau4} \mid}^2   \\
{\mid {\tilde{U}}_{s1} \mid}^2 & {\mid {\tilde{U}}_{s2} \mid}^2 &
{\mid {\tilde{U}}_{s3} \mid}^2 & {\mid {\tilde{U}}_{s4} \mid}^2
\end{array} \right)
\left (\begin{array}{cccc}
{\mid {\tilde{U}}_{e1} \mid}^2 & {\mid {\tilde{U}}_{\mu1} \mid}^2 &
{\mid {\tilde{U}}_{\tau1} \mid}^2 & {\mid {\tilde{U}}_{s1} \mid}^2  \\
{\mid {\tilde{U}}_{e2} \mid}^2 & {\mid {\tilde{U}}_{\mu2} \mid}^2 &
{\mid {\tilde{U}}_{\tau2} \mid}^2 & {\mid {\tilde{U}}_{s2} \mid}^2  \\
{\mid {\tilde{U}}_{e3} \mid}^2 & {\mid {\tilde{U}}_{\mu3} \mid}^2 &
{\mid {\tilde{U}}_{\tau3} \mid}^2 & {\mid {\tilde{U}}_{s3} \mid}^2  \\
{\mid {\tilde{U}}_{e4} \mid}^2 & {\mid {\tilde{U}}_{\mu4} \mid}^2 &
{\mid {\tilde{U}}_{\tau4} \mid}^2 & {\mid {\tilde{U}}_{s4} \mid}^2
\end{array} \right)\nonumber \\
& &  \times
\left (\begin{array}{c}
1 \\
2 \\
0 \\
0 \end{array} \right )
\phi_{\nu_e}\,\, .
\label{cosflux1}
\eea
}
Eq. (\ref{cosflux1}) follows that
{\small
\bea
F^4_{\nu_e} &=& [ {\mid {\tilde{U}}_{e1} \mid}^2 (1 + {\mid {\tilde{U}}_{\mu1} \mid}^2 - {\mid {\tilde{U}}_{\tau1} \mid}^2  - {\mid {\tilde{U}}_{s1} \mid}^2 )
+ {\mid {\tilde{U}}_{e2} \mid}^2 (1 + {\mid {\tilde{U}}_{\mu2} \mid}^2  -
{\mid {\tilde{U}}_{\tau2} \mid}^2  - {\mid {\tilde{U}}_{s2} \mid}^2 )\nonumber \\& &  + {\mid {\tilde{U}}_{e3} \mid}^2 (1 + {\mid {\tilde{U}}_{\mu3} \mid}^2 -
{\mid {\tilde{U}}_{\tau3} \mid}^2 - {\mid {\tilde{U}}_{s3} \mid}^2 )
+ {\mid {\tilde{U}}_{e4} \mid}^2 (1 + {\mid {\tilde{U}}_{\mu4} \mid}^2 -
{\mid {\tilde{U}}_{\tau4} \mid}^2 - {\mid {\tilde{U}}_{s4} \mid}^2 )]\phi_{\nu_e}\,\,\, ,\nonumber\\
F^4_{\nu_\mu} &=& [ {\mid {\tilde{U}}_{\mu1} \mid}^2 (1 + {\mid {\tilde{U}}_{\mu1} \mid}^2 - {\mid {\tilde{U}}_{\tau1} \mid}^2  - {\mid {\tilde{U}}_{s1} \mid}^2 ) + {\mid {\tilde{U}}_{\mu2} \mid}^2 (1 + {\mid {\tilde{U}}_{\mu2} \mid}^2  -
{\mid {\tilde{U}}_{\tau2} \mid}^2  - {\mid {\tilde{U}}_{s2} \mid}^2 )\nonumber\\
& &+ {\mid {\tilde{U}}_{\mu3} \mid}^2 (1 + {\mid {\tilde{U}}_{\mu3} \mid}^2 -
{\mid {\tilde{U}}_{\tau3} \mid}^2 - {\mid {\tilde{U}}_{s3} \mid}^2 )
+ {\mid {\tilde{U}}_{\mu4} \mid}^2 (1 + {\mid {\tilde{U}}_{\mu4} \mid}^2 -
{\mid {\tilde{U}}_{\tau4} \mid}^2 - {\mid {\tilde{U}}_{s4} \mid}^2 )]\phi_{\nu_e}\,\,\, ,\nonumber\\
F^4_{\nu_\tau} &=& [ {\mid {\tilde{U}}_{\tau1} \mid}^2 (1 + {\mid {\tilde{U}}_{\mu1} \mid}^2 - {\mid {\tilde{U}}_{\tau1} \mid}^2  - {\mid {\tilde{U}}_{s1} \mid}^2 ) + {\mid {\tilde{U}}_{\tau2} \mid}^2 (1 + {\mid {\tilde{U}}_{\mu2} \mid}^2  - {\mid {\tilde{U}}_{\tau2} \mid}^2  - {\mid {\tilde{U}}_{s2} \mid}^2 )\nonumber\\& & + {\mid {\tilde{U}}_{\tau3} \mid}^2 (1 + {\mid {\tilde{U}}_{\mu3} \mid}^2 -{\mid {\tilde{U}}_{\tau3} \mid}^2 - {\mid {\tilde{U}}_{s3} \mid}^2 )
+ {\mid {\tilde{U}}_{\tau4} \mid}^2 (1 + {\mid {\tilde{U}}_{\mu4} \mid}^2 -
{\mid {\tilde{U}}_{\tau4} \mid}^2 - {\mid {\tilde{U}}_{s4} \mid}^2 )]\phi_{\nu_e}\,\,\, ,\nonumber\\
F^4_{\nu_s} &=& [ {\mid {\tilde{U}}_{s1} \mid}^2 (1 + {\mid {\tilde{U}}_{\mu1} \mid}^2 - {\mid {\tilde{U}}_{\tau1} \mid}^2  - {\mid {\tilde{U}}_{s1} \mid}^2 )
+ {\mid {\tilde{U}}_{s2} \mid}^2 (1 + {\mid {\tilde{U}}_{\mu2} \mid}^2  -
{\mid {\tilde{U}}_{\tau2} \mid}^2  - {\mid {\tilde{U}}_{s2} \mid}^2 )\nonumber\\
& & + {\mid {\tilde{U}}_{s3} \mid}^2 (1 + {\mid {\tilde{U}}_{\mu3} \mid}^2 -
{\mid {\tilde{U}}_{\tau3} \mid}^2 - {\mid {\tilde{U}}_{s3} \mid}^2 )\,\,\nonumber\\
& &+ {\mid {\tilde{U}}_{s4} \mid}^2 (1 + {\mid {\tilde{U}}_{\mu4} \mid}^2 -
{\mid {\tilde{U}}_{\tau4} \mid}^2 - {\mid {\tilde{U}}_{s4} \mid}^2 )]\phi_{\nu_e}\,\,\, ,
\label{4fmuprob}
\eea
}
where 
{\small
\bea
\tilde{U} &=& \left (\begin{array}{cccc}
c_{14}{\cal{U}}_{e1} & c_{14}{\cal{U}}_{e2} & c_{14}{\cal{U}}_{e3} & s_{14}  \\
& & & \\
-s_{14}s_{24}{\cal{U}}_{e1}+c_{24}{\cal{U}}_{\mu1} &
-s_{14}s_{24}{\cal{U}}_{e2}+c_{24}{\cal{U}}_{\mu2} &
-s_{14}s_{24}{\cal{U}}_{e3}+c_{24}{\cal{U}}_{\mu3} & c_{14}s_{24}  \\
&&& \\
\begin{array}{c}
-c_{24}s_{14}s_{34}{\cal{U}}_{e1}\\
-s{24}s{34}{\cal{U}}_{\mu1}\\
+c_{34}{\cal{U}}_{\tau1} \end{array} &
\begin{array}{c}
-c_{24}s_{14}s_{34}{\cal{U}}_{e2}\\
-s{24}s{34}{\cal{U}}_{\mu2}\\
+c_{34}{\cal{U}}_{\tau2} \end{array}  &
\begin{array}{c}
-c_{24}s_{14}s_{34}{\cal{U}}_{e3}\\
-s{24}s{34}{\cal{U}}_{\mu3}\\
+c_{34}{\cal{U}}_{\tau3} \end{array}  &
c_{14}c_{24}s_{34}    \\
&&& \\
\begin{array}{c}
-c_{24}c_{34}s_{14}{\cal{U}}_{e1}\\
-s_{24}c_{34}{\cal{U}}_{\mu1}\\
-s_{34}{\cal{U}}_{\tau1} \end{array}  &
\begin{array}{c}
-c_{24}c_{34}s_{14}{\cal{U}}_{e2}\\
-s_{24}c_{34}{\cal{U}}_{\mu2}\\
-s_{34}{\cal{U}}_{\tau2} \end{array}  &
\begin{array}{c}
-c_{24}c_{34}s_{14}{\cal{U}}_{e3}\\
-s_{24}c_{34}{\cal{U}}_{\mu3}\\
-s_{34}{\cal{U}}_{\tau3} \end{array}  &
c_{14}c_{24}c_{34}  \end{array} \right )\,\,\,\,\,\,\,\,.
\label{pmns4}
\eea
}
In our calculation, we consider that the intrinsic electron neutrino flux at
the source ($\phi_{\nu_e}$) is equivalent to the theoretical flux
$\phi^{\rm th} (E_\nu)$ (mentioned in the Section 2)
obtained from the decay of superheavy dark matter particles. From the above 
formalism it is evident that the computation of neutrino flux of any 
flavour on reaching the Earth (after undergoing neutrino oscillation) can 
be done if $\nu_e$ flux at the source can be computed and 
with the proper evaluation of the PMNS mixing matrix elements.
\section{Calculations and Results}
We propose in this work that the high energy neutrino events detected by 
IceCube could have originated from the decay of superheavy dark matter. 
In case some events might have astrophysical origin, we include in our 
analyses, this possibility also. By these analyses, we 
demonstrate that while the two reported events in the energy range $\sim$ 60 
TeV - $\sim$ 120 TeV can be best explained when the astrophysical flux is 
considered in 
the analysis and the rest can be very well fitted with both the hadronic and 
leptonic channel production of SHDM decay. 
\subsection{The choice and data}
We have considered the energy region $\sim$ 60 TeV - $\sim 5 \times 10^7$ GeV 
in our analysis. The 
event data points as given by IceCube Collaboration 
(Fig. 2 of Ref. \cite{icrc}) can be categorized in three regions 
for three energy ranges. In Fig. 1, we have reproduced, from Fig. 2 of 
Ref. \cite{icrc}, the energy range (and event data points) from which 
the data sets for the present analysis has been chosen. 

\noindent{1. \underline{Range $\sim$ 60 TeV to $\sim$ 120 TeV}}\\
There are two event points in the range $\sim$ 60 TeV-$\sim$ 120 TeV. These 
two points are adopted in our analysis.

\noindent{2. \underline{Range $\sim 1.2 \times 10^5$ GeV - $\sim$ 5 $\times 10^6$ GeV}}\\
There are four data points in the energy region $\sim 1.2 \times 10^5$ GeV to 
$\sim 5 \times 10^6$ GeV. For one of the four points, only upper limit 
is given. This region is designated by a pink band 
(Fig. 2 of Ref. \cite{icrc} and Fig. 1) and is used for the analysis of the 
upgoing muon neutrino spectrum in this region. The width of the band 
indicates 1$\sigma$ uncertainty. In our analysis (with dark matter decay 
consideration) we adopt the three points in this region (along with the errors) 
that is included in the pink band and also choose another 12 points from 
the pink band with the error given by the 
width of the band. Thus, in this region we have a total of 15 points.

\noindent{3. \underline{Range $\sim$ 5 $\times 10^6$ GeV - $\sim$ 
5 $\times 10^7$ GeV}}\\
In this energy range, the IC Collaboration in Fig. 2 of Ref. \cite{icrc}, 
indicates upper limits of four event points. One may immediately notice that 
the nature of these four upper bounds grossly differ from that of the pink 
band. In this work we adopt these upper bounds as event points. 

All the 21 event points (henceforth referred to as ``data points") used in the 
present analyses are enlisted in Table 1.

\subsection{Definition of $\chi^2$}
We have made $\chi^2$ analyses with data points of the whole range of energies 
(or for two different ranges together) to understand the role of different 
components of our proposed dark matter decay origin (through either or both 
hadronic and leptonic channels) of ultra high energy neutrinos 
as well as the astrophysical components and 
for interpreting the events in case of three different 
regions mentioned above. The parameters in the analysis are the mass of 
superheavy dark matter $m_{\chi}$ and the decay lifetime $\tau$ which are 
obtained from $\chi^2$ fitting of the data points.

The $\chi^2$ for our analysis is defined as 
\bea
\chi^2 &=& \sum_{i = 1}^{n} \left (\displaystyle\frac {E_i^2 \phi_i^{\rm th} -
E_i^2 \phi_i^{\rm Ex}} {(\rm err)_i} \right )^2 \,\, ,
\label{cal1}
\eea
where $n$ is the number of data points. Note that $n=21$ when the whole 
range of energy is considered, $n=17$ if only energy ranges 1 and 2 are 
considered etc. In the above,  $\phi_i^{\rm th}$ designates the theoretical 
flux. For the total energy range, therefore the total flux is as given in 
Eq. (\ref{form5}), while in case the analysis is performed with partial 
energy ranges, relevant fluxes or their sum will be considered. In 
Eq. (\ref{cal1}),  $\phi_i^{\rm Ex}$ denotes the experimental data point at 
energy $E_i$ with error (err)$_i$.   

For the $\chi^2$ fit, the theoretical fluxes for electron neutrinos are then 
computed using Eqs. (\ref{flux})-(\ref{form5}). After the neutrinos undergo
oscillation on reaching the Earth, the muon neutrino flux on arrival is 
obtained from Eq. (\ref{3fmuprob}) (for 3-flavour case) or Eq. (\ref{4fmuprob}) 
(for 4-flavour case).

\begin{table}[]
\centering
\caption{The selected data points. see text for details. (The error bars
for the last four data points are chosen to be the values of data points
itself as there are no lower bounds for those data points}
\vskip 2mm
\label{1}
\begin{tabular}{cccc}
\hline
Energy  & Neutrino Flux ($E_\nu^2 \displaystyle\frac {d\Phi} {dE}$)  & \multirow{2}{*}{Error}
\\
(in GeV) & (in GeV cm$^{-2}$ s$^{-1}$ sr$^{-1}$) & \\ \hline
\hline
6.13446$\times 10^4\, ^*$ & 2.23637$\times 10^8$ & 2.16107$\times 10^8$\\
1.27832$\times 10^5\, ^*$ & 2.70154$\times 10^8$ & 1.30356$\times 10^8$\\
\hline
2.69271$\times 10^5\, ^*$ & 7.66476$\times 10^9$ & 8.5082$\times 10^9$\\
1.19479$\times 10^6\, ^*$ & 5.14335$\times 10^9$ & 7.6982$\times 10^9$\\
2.51676$\times 10^6\, ^*$ & 4.34808$\times 10^9$ & 8.4481$\times 10^9$\\
3.54813$\times 10^6$ & 5.25248$\times 10^9$ & 4.1258$\times 10^9$\\
2.30409$\times 10^6$ & 5.71267$\times 10^9$ & 4.1600$\times 10^9$\\
1.52889$\times 10^6$ & 6.21317$\times 10^9$ & 3.9882$\times 10^9$\\
1.05925$\times 10^6$ & 6.61712$\times 10^9$ & 3.7349$\times 10^9$\\
7.18208$\times 10^5$ & 7.04733$\times 10^9$ & 3.9777$\times 10^9$\\
4.46684$\times 10^5$ & 7.66476$\times 10^9$ & 3.6478$\times 10^9$\\
2.86954$\times 10^5$ & 8.16308$\times 10^9$ & 4.1571$\times 10^9$\\
1.90409$\times 10^5$ & 8.87827$\times 10^9$ & 6.2069$\times 10^9$\\
1.43818$\times 10^5$ & 9.65612$\times 10^9$ & 6.8856$\times 10^9$\\
2.51189$\times 10^6$ & 4.16928$\times 10^9$ & 8.2726$\times 10^9$\\
1.19279$\times 10^6$ & 5.03649$\times 10^9$ & 7.5383$\times 10^9$\\
2.68960$\times 10^5$ & 7.50551$\times 10^9$ & 8.1583$\times 10^9$\\
\hline
5.30143$\times 10^6\, ^*$ & 1.55414$\times 10^9$ & \\
1.10473$\times 10^7\, ^*$ & 4.08265$\times 10^9$ & \\
2.32705$\times 10^7\, ^*$ & 6.08407$\times 10^9$ & \\
4.90181$\times 10^7\, ^*$ & 1.05021$\times 10^8$ & \\
\hline
\end{tabular}
\end{table}

\subsection{Analysis}
Following the formalism described above and Section 3 we make the $\chi^2$ fit 
by $\chi^2$ minimisation with the data given in Table 1. The use has been 
made of Eqs. (\ref{flux}) - (\ref{pmns4}) for computations of theoretical flux 
components namely astrophysical flux and those predicted from our proposition 
of the decay of a superheavy dark matter via the hadronic and leptonic 
channels. From the fit, the best fit values of the two unknown 
parameters of the formalism namely the mass $m_{\chi}$ and the lifetime 
$\tau$ of the decaying dark matter are obtained. The $1\sigma, 2\sigma$ and 
$3\sigma$ ranges for each of the $\chi^2$ analysis are also computed and 
furnished along with the study of the quantity of fit for different 
chosen data sets from Table 1.

We furnish the analyses by considering six different cases. These are given 
below.
\begin{itemize}
\item{Case I} - All 21 points of the Table 1 is fitted with the theoretical 
flux 
at source as in Eq. (\ref{form5}) where astrophysical flux at source computed 
as in Eq. (\ref{flux}).
\item{Case II} - Same as Case I but for astrophysical source flux computed 
as in 
Eq. (\ref{icflux}) (the power law spectrum given by IC Collaboration).
\item{Case III} - All 21 points of Table 1. But the theoretical flux is 
computed without the astrphysical component (only the hadronic and leptonic 
channel of dark matter decay)
$$
\phi^{\rm th} (E_\nu) = \left (\displaystyle\frac {d\Phi_{\rm EG}} {dE} (E_\nu) \right)_{\rm had} +
\left (\displaystyle\frac {d\Phi_{\rm G}} {dE} (E_\nu) \right)_{\rm had} +
\left (\displaystyle\frac {d\Phi_{\rm EG}} {dE} (E_\nu) \right)_{\rm lep} +
\left (\displaystyle\frac {d\Phi_{\rm G}} {dE} (E_\nu) \right)_{\rm lep} \,\, .
$$
\item{Case IV}- All 21 points of Table 1. Theoretical flux 
$$
\phi^{\rm th} (E_\nu) =  \displaystyle\frac {d \phi_{\nu_{\rm Ast}}}
{d E_\nu} (E_\nu) +
\left (\displaystyle\frac {d\Phi_{\rm EG}} {dE} (E_\nu) \right)_{\rm had} +
\left (\displaystyle\frac {d\Phi_{\rm G}} {dE} (E_\nu) \right)_{\rm had} \,\, 
$$
(no leptonic channel for dark matter decay, the astrophysical flux is from 
Eq. (\ref{flux})).
\item{Case V}- The last four points of Table 1 excluded. Total number of 
points = 17. The theoretical flux is calculated as 
$$
\phi^{\rm th} (E_\nu) =  \displaystyle\frac {d \phi_{\nu_{\rm Ast}}}
{d E_\nu} (E_\nu) +
\left (\displaystyle\frac {d\Phi_{\rm EG}} {dE} (E_\nu) \right)_{\rm had} +
\left (\displaystyle\frac {d\Phi_{\rm G}} {dE} (E_\nu) \right)_{\rm had} \,\, 
$$
(no leptonic decay channel and the theroretical astrophysical flux is 
from Eq. (\ref{flux})).
%\item{Case VI}- First two data points of Table 1 are not taken. Total number 
%of points = 19. Theoretical flux as in Eq. (\ref{form5}) with 
%Eq. (\ref{icflux}) for theoretical astrophysical flux.
%\item{Case VII}- Same as Case VI but for theoretical astrophysical flux 
%Eq. (\ref{flux}) is employed.
\end{itemize} 

In Figs. 2 - 6 we show (a) the fluxes calculated with the fitted values using 
corresponding theoretical flux formula and (b) the $m_{\chi}$-$\tau$ contour 
plot 
with $1\sigma, 2\sigma$ and $3\sigma$ contours. The best fit values for 
$m_{\chi}$ and $\tau$ are also shown.

\begin{itemize}
\item It is shown from Figs. 2 - 3 that the first two points of Table 1 (in the 
energy range $\sim$ 60 TeV - $\sim$ 120 TeV) cannot be fitted if astrophysical 
flux is not considered.
\item The data points in the energy range ($\sim 1.2 \times 10^5$ GeV - 
$\sim 5 \times 10^7$ GeV) 
can be well explained from the consideration that these neutrinos originate 
from the decay of superheavy dark matter.
\item The neutrino events within the energy range 
$\sim 1.2 \times 10^5$ GeV - $\sim 5 \times 10^6$ GeV (pink band) is very well 
represented by the 
neutrinos produced from the SHDM decay via {\it hadronic channel}.
\item Fig. 5(a) and Fig. 6(a) clearly demonstrate that only hadronic channel 
cannot explain the events beyond the energy $\sim 5 \times 10^6$ GeV - 
$\sim 5 \times 10^7$ GeV. There are indications from this analysis that 
the data 
events in the energy range $\sim 5 \times 10^6$ GeV - $\sim 5 \times 10^7$ GeV 
can only 
be represented by the neutrinos from {\it leptonic channel} of SHDM decay 
in the present framework.
\end{itemize}

\begin{table}[]
        \centering
        \caption{Best fit values of $m_\chi$ and $\tau$ for different 
cases.}
        \vskip 2mm
        \label{1}
        \begin{tabular}{cccc}
                \hline
                \multirow{2}{*}{Set} & $m_{\chi}$ & $\tau$ & Value of $\chi_{min}^2$ \\
                & in GeV & in sec &\\
                \hline \hline
                All points, All channels & \multirow{2}{*}{1.5461$\times 10^8$} & \multirow{2}{*}{2.2136$\times 10^{29}$} & \multirow{2}{*}{3.8744}\\
                (the astrophysical flux &&&\\ 
                is taken from Eq. (\ref{flux})) &&&\\ \hline
                All points, All channels & \multirow{2}{*}{1.4765$\times 10^8$} & \multirow{2}{*}{5.6898$\times 10^{29}$} & \multirow{2}{*}{9.5042}\\
                (the flux adopted from the &&&\\
                IC Collaboration (Eq. (\ref{icflux})) as &&&\\
                the astrophysical flux) &&&\\ \hline
                All points, Leptonic channel,  & \multirow{2}{*}{1.5640$\times 10^8$} & \multirow{2}{*}{1.6410$\times 10^{29}$} & \multirow{2}{*}{9.5208}\\
                Hadronic channel &&&\\ \hline
                All points, Hadronic channel  & \multirow{2}{*}{7.586$\times 10^7$} & \multirow{2}{*}{1.603$\times 10^{29}$} & \multirow{2}{*}{4.6504}\\
                (pink band points), &&&\\
                 Astrophysical flux (Eq. (\ref{flux})) &&&\\ \hline
                First 5 points, Hadronic channel  & \multirow{2}{*}{1.2679$\times 10^8$} & \multirow{2}{*}{1.5314$\times 10^{29}$} & \multirow{2}{*}{1.5301}\\
                (pink band points), &&&\\
                Astrophysical flux (Eq. (\ref{flux})) &&&\\ \hline
%
%                No astro points, & \multirow{2}{*}{1.4765$\times 10^8$} & \multirow{2}{*}{5.6898$\times 10^{29}$} & \multirow{2}{*}{8.4332}\\
%                All channels (IceCube) &&&\\ \hline
%                No astro points, & \multirow{2}{*}{1.5461$\times 10^8$} & \multirow{2}{*}{2.2652$\times 10^{29}$} & \multirow{2}{*}{2.4919}\\
%                All channels (Chianese) &&&\\ \hline
%                No astro points, & \multirow{2}{*}{1.5640$\times 10^8$} & \multirow{2}{*}{1.6792$\times 10^{29}$} & \multirow{2}{*}{5.6283}\\
%                Leptonic+Hadronic channel &&&\\ \hline
               \hline
        \end{tabular}
\end{table}
As mentioned, the best fit values of the parameters $m_{\chi}$ and $\tau$ 
(as well as $1\sigma, 2\sigma$ and $3\sigma$ contours) for each of the cases 
are shown in Fig. 2(b) - 6(b) respectively. These values along with the 
respective values for $\chi^2_{\rm min}$ are shown in Table 2. 

From the above analyses therefore, it can be stated that the UHE neutrino 
events reported by IC in the energy range $\sim 1.2 \times 10^5$ GeV to 
$\sim 5 \times 10^7$ GeV can be well described to have originated from 
the decay of a supermassive dark matter of mass $\sim 10^8$ GeV and decay 
time $\sim 10^{29}$ sec. 

We also like to state that we repeat the entire analysis with 4-flavour 
oscillation scenario with no or any significant changes. For this purpose 
the values of mixing angles are chosen as $\theta_{14} = 3.6^o , 
\theta_{24} = 4^o, \theta_{34} = 18.48^o$. These values are 
within the allowed limits of the analyses of 
NOvA \cite{nova}, MINOS \cite{minos}, Daya Bay \cite{daya} 
neutrino experiments.

In Fig. 7 we furnish a ternary plot showing the flavour ratios of the active 
neutrinos on reaching the Earth for both the 3-flavour and 4-flavour 
oscillation cases. The flavour ($\nu_e,\nu_{\mu},\nu_{\tau}$) ratio of 1:1:1 
for active neutrinos are also shown for comparison. The assumed production 
ratio of 1:2:0 for the active neutrinos is also furnished. The flavour ratio 
barely changes when 4-flavour oscillation is considered.

\begin{figure}[h!]
\centering
{\includegraphics[height=6.0 cm,angle=0]{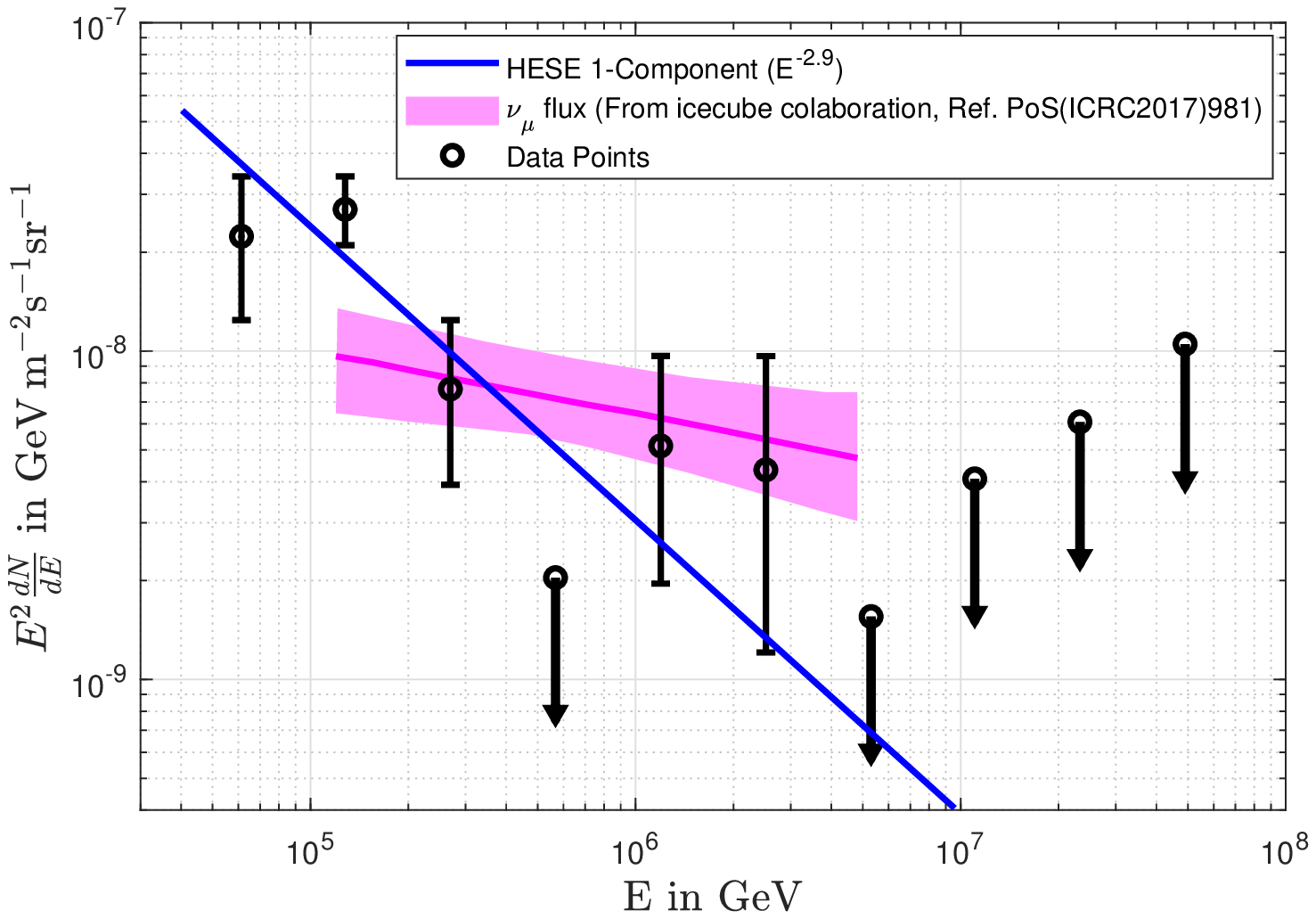}}
\caption{The data points and range of energy considered in the present 
analyses. The pink band is also shown. These are reproduced from Fig. 2 
of Ref. \cite{icrc}}
\label{fig1}
\end{figure}

\begin{figure}[h!]
\centering
\subfigure[]{
\includegraphics[height=6.0 cm, width=6.0 cm,angle=0]{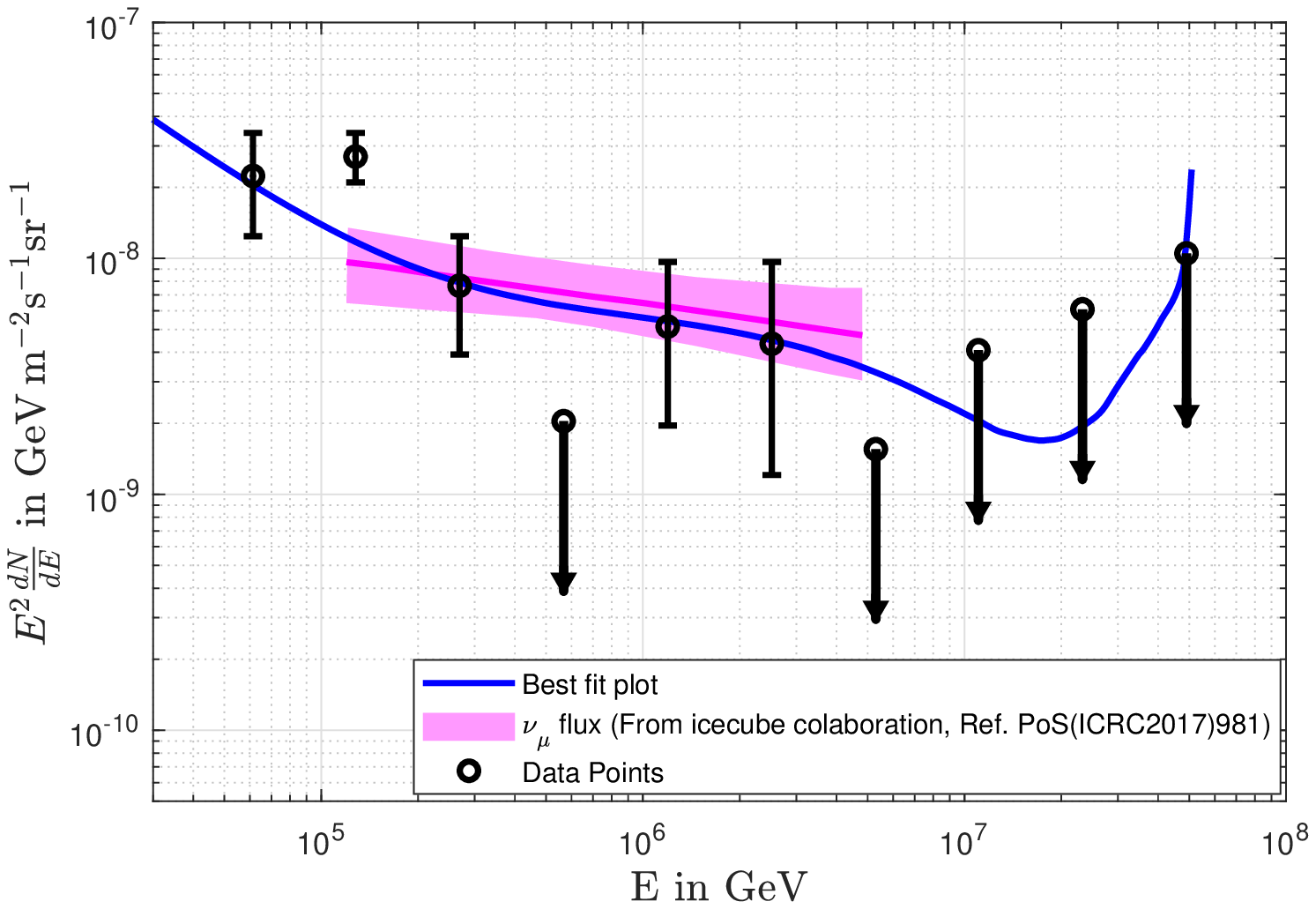}}
\subfigure []{
\includegraphics[height=6.0 cm, width=6.0 cm,angle=0]{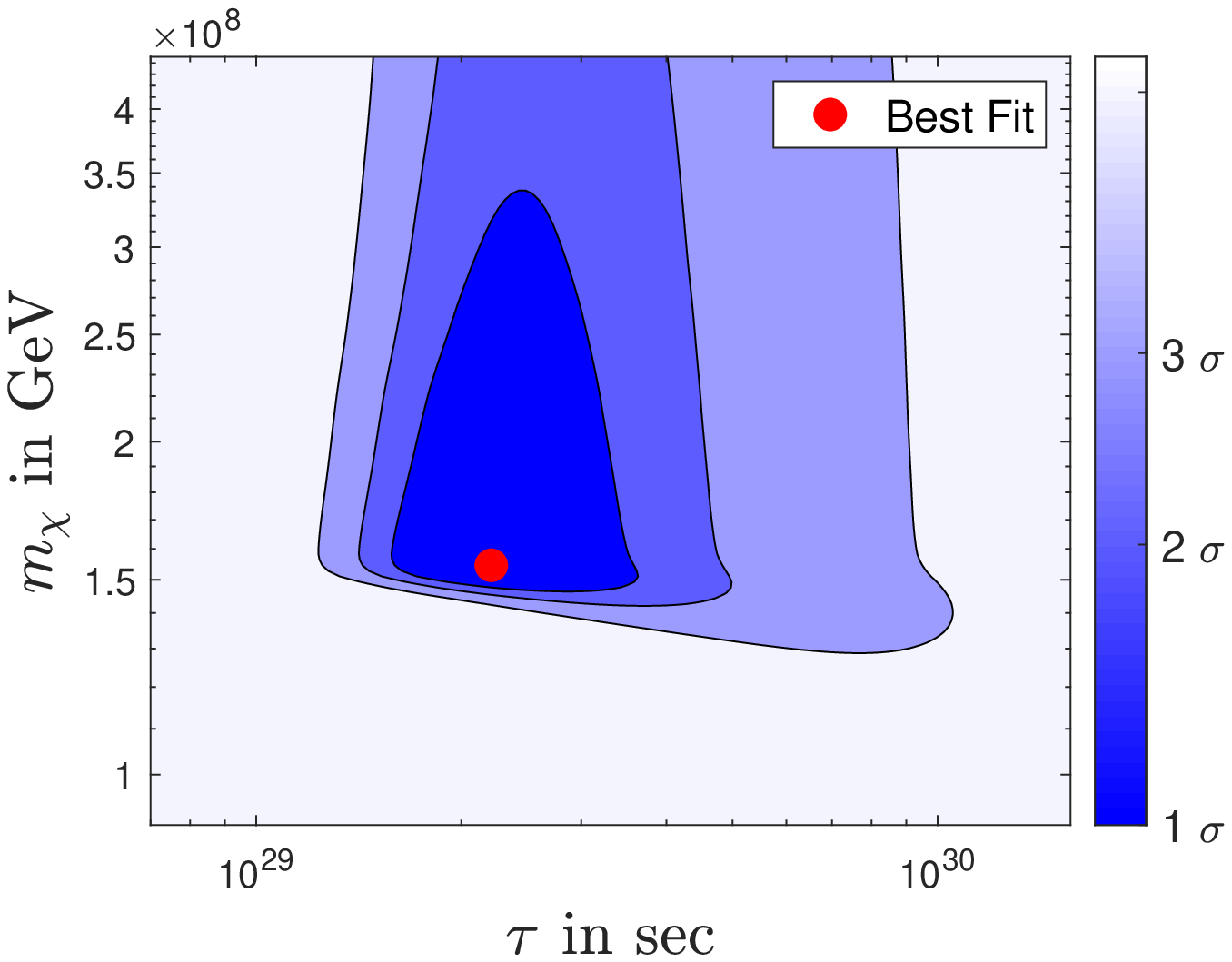}}
\caption{(a)flux, (b) contour by considering all points, all channels
(The astrophysical flux is computed from Eq. (\ref{flux})). See text for details.}
\label{figr2}
\end{figure}

\begin{figure}[h!]
\centering
\subfigure[]{
\includegraphics[height=6.0 cm, width=6.0 cm,angle=0]{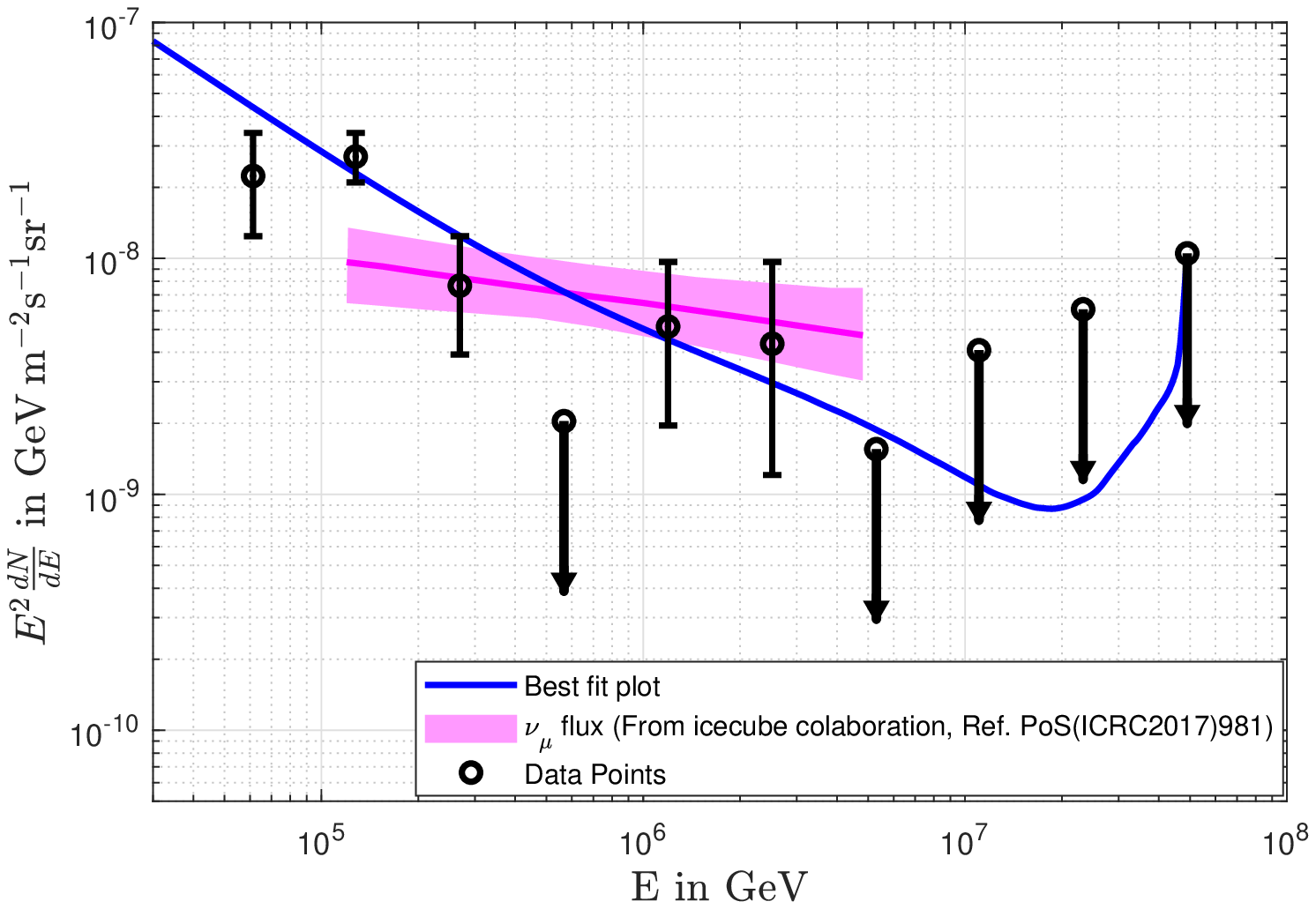}}
\subfigure []{
\includegraphics[height=6.0 cm, width=6.0 cm,angle=0]{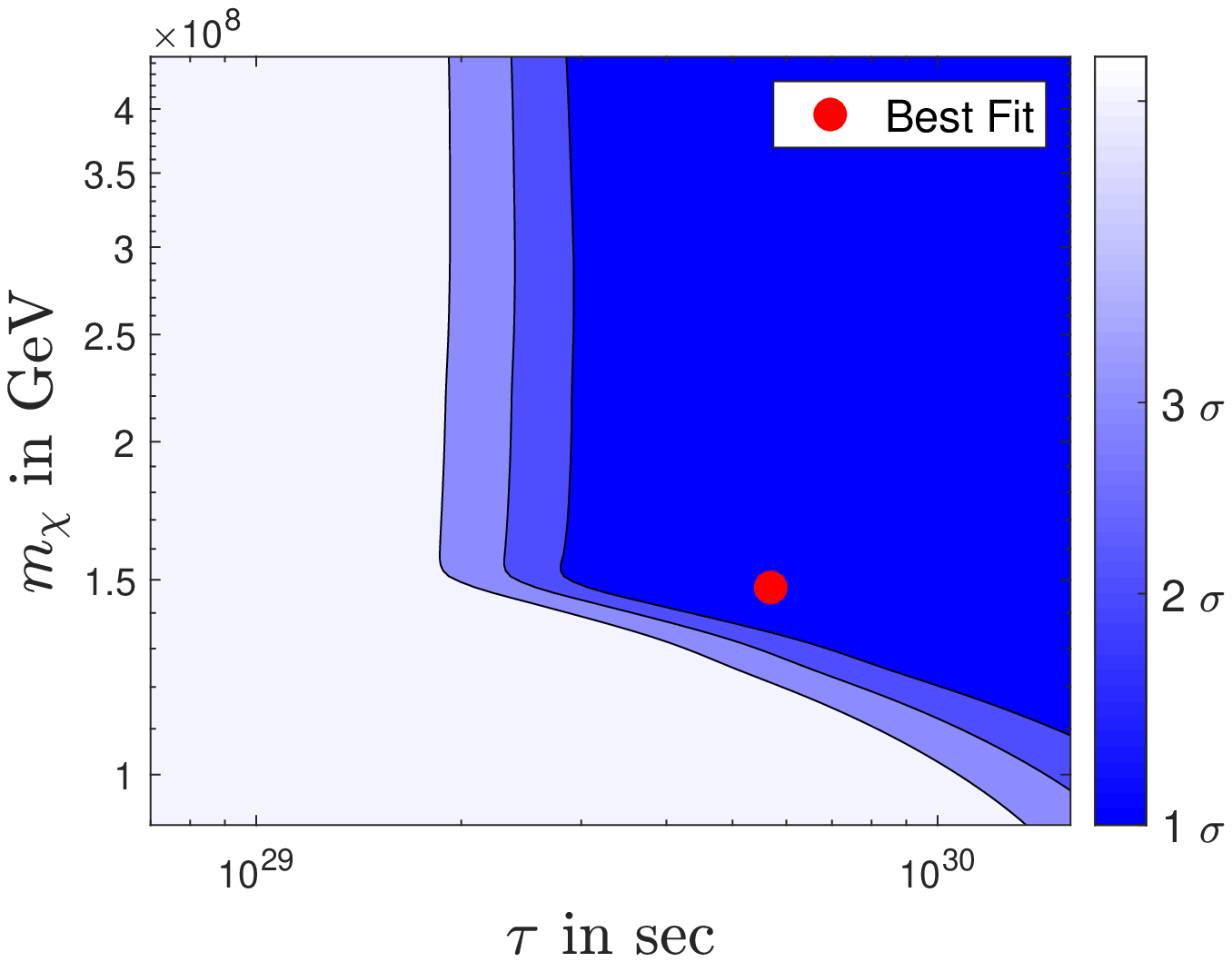}}
\caption{(a)flux, (b) contour by considering all points, all channels
(the flux given by the IC Collaboration is considered as the 
astrophysical flux (Eq. (\ref{icflux}))). See text for details.}
\label{figr3}
\end{figure}

\begin{figure}[h!]
\centering
\subfigure[]{
\includegraphics[height=6.0 cm, width=6.0 cm,angle=0]{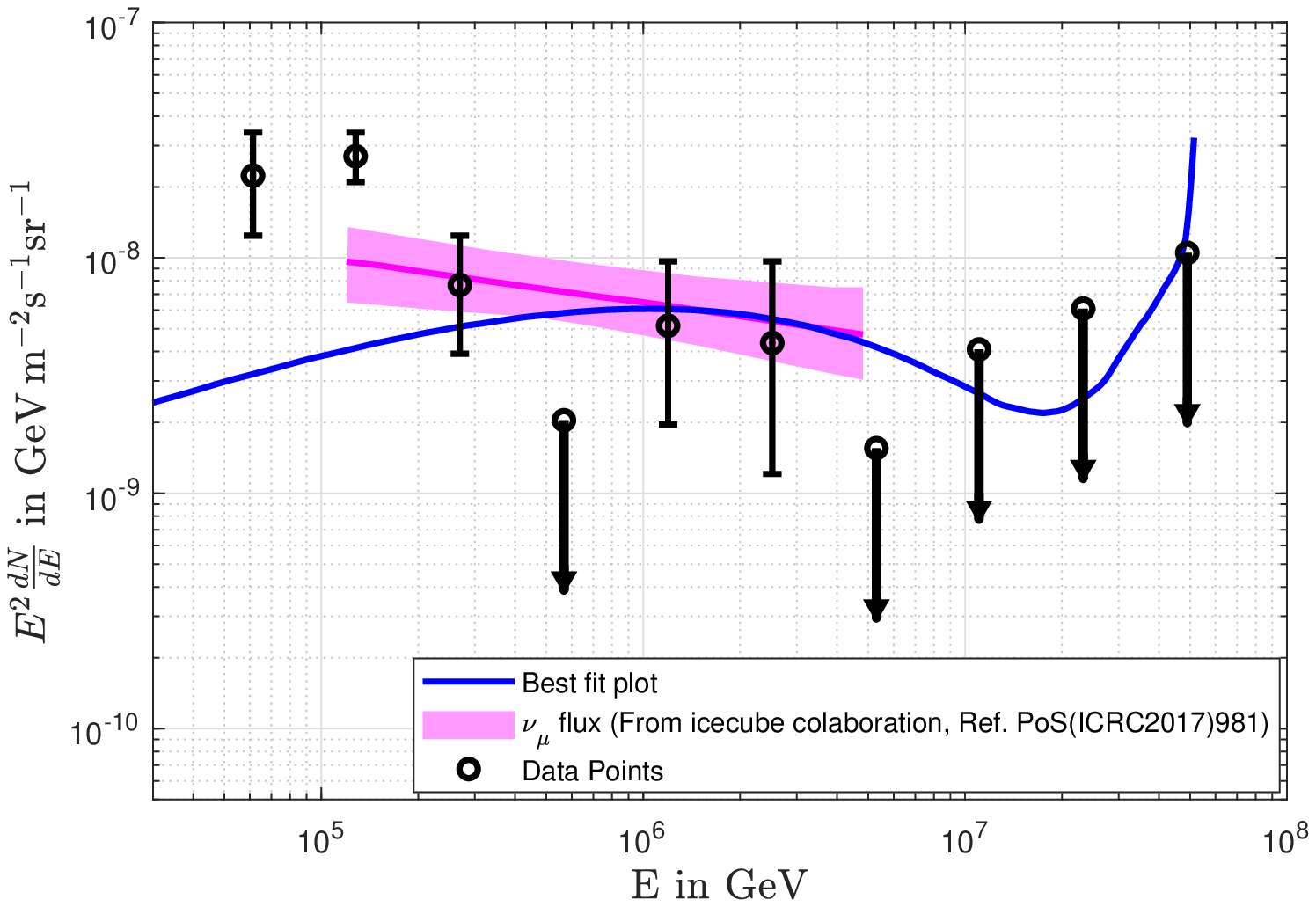}}
\subfigure []{
\includegraphics[height=6.0 cm, width=6.0 cm,angle=0]{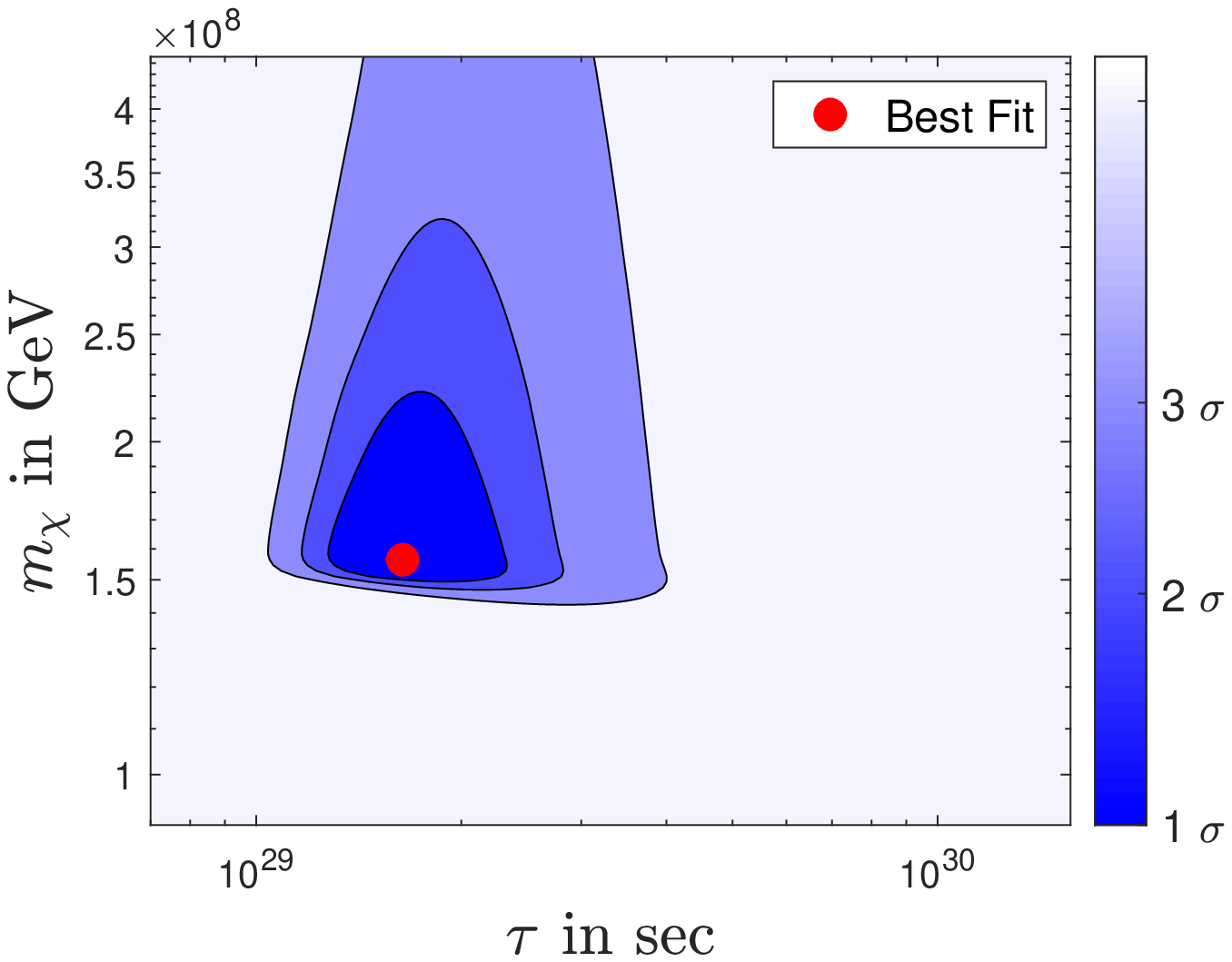}}
\caption{(a)flux, (b) contour by considering all points, leptonoic+hadronic
channels . See text for details.}
\label{figr4}
\end{figure}

\begin{figure}[h!]
\centering
\subfigure[]{
\includegraphics[height=6.0 cm, width=6.0 cm,angle=0]{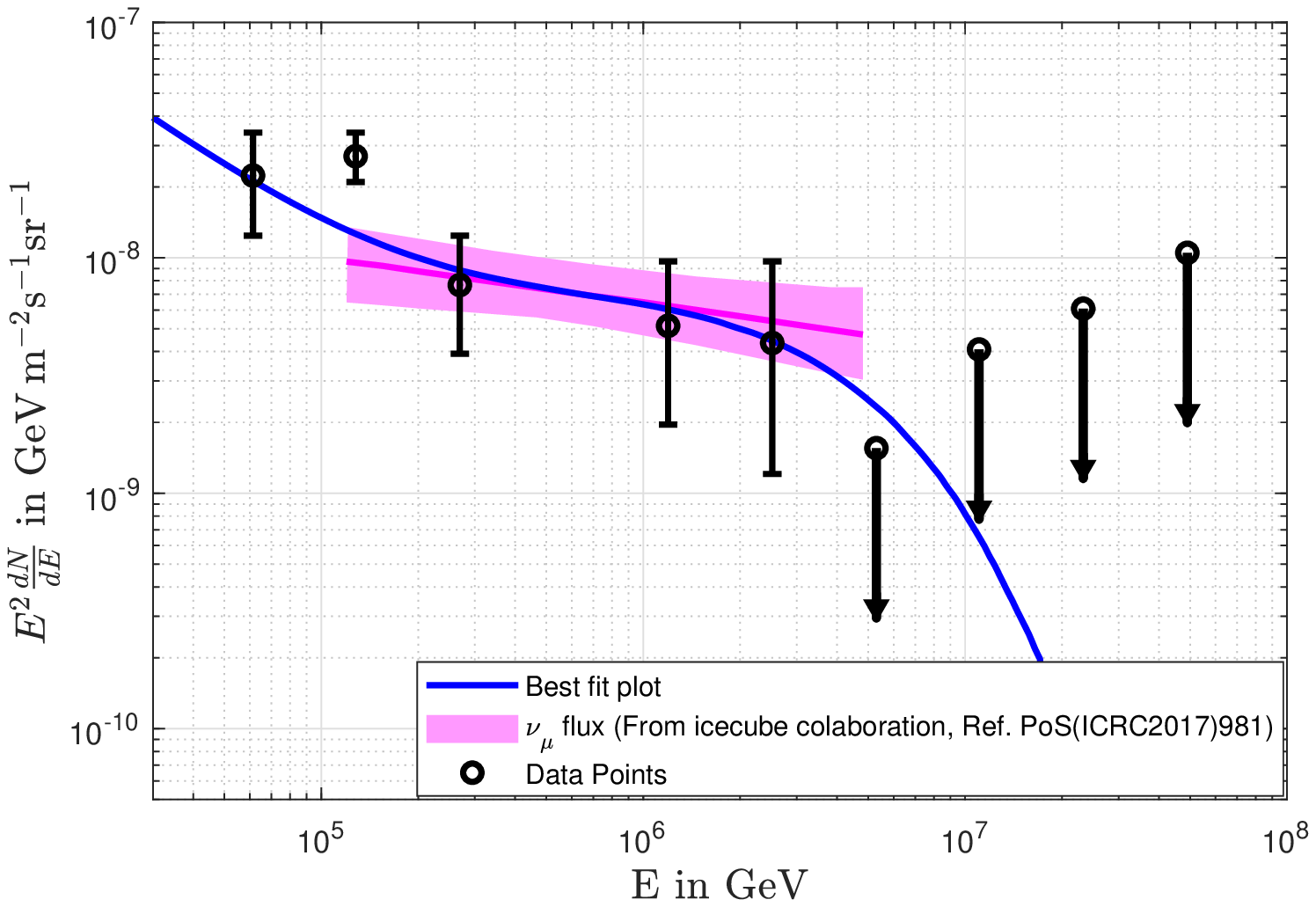}}
\subfigure []{
\includegraphics[height=6.0 cm, width=6.0 cm,angle=0]{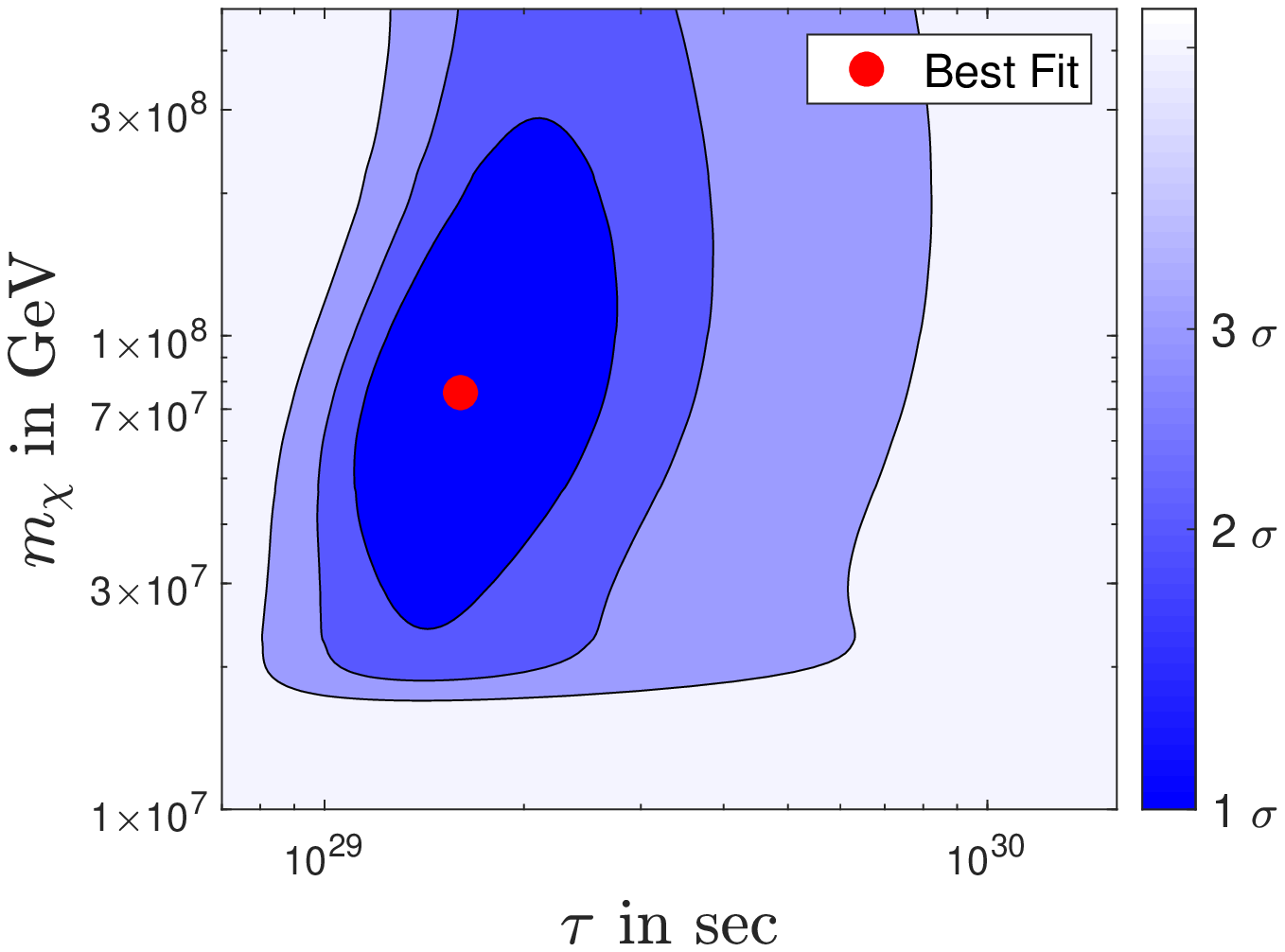}}
\caption{(a)flux, (b) contour by considering all points, 
hadronic+astro channels
(the astrophysical flux is computed from Eq. (\ref{flux})). 
See text for details.}
\label{figr5}
\end{figure}

\begin{figure}[h!]
\centering
\subfigure[]{
\includegraphics[height=6.0 cm, width=6.0 cm,angle=0]{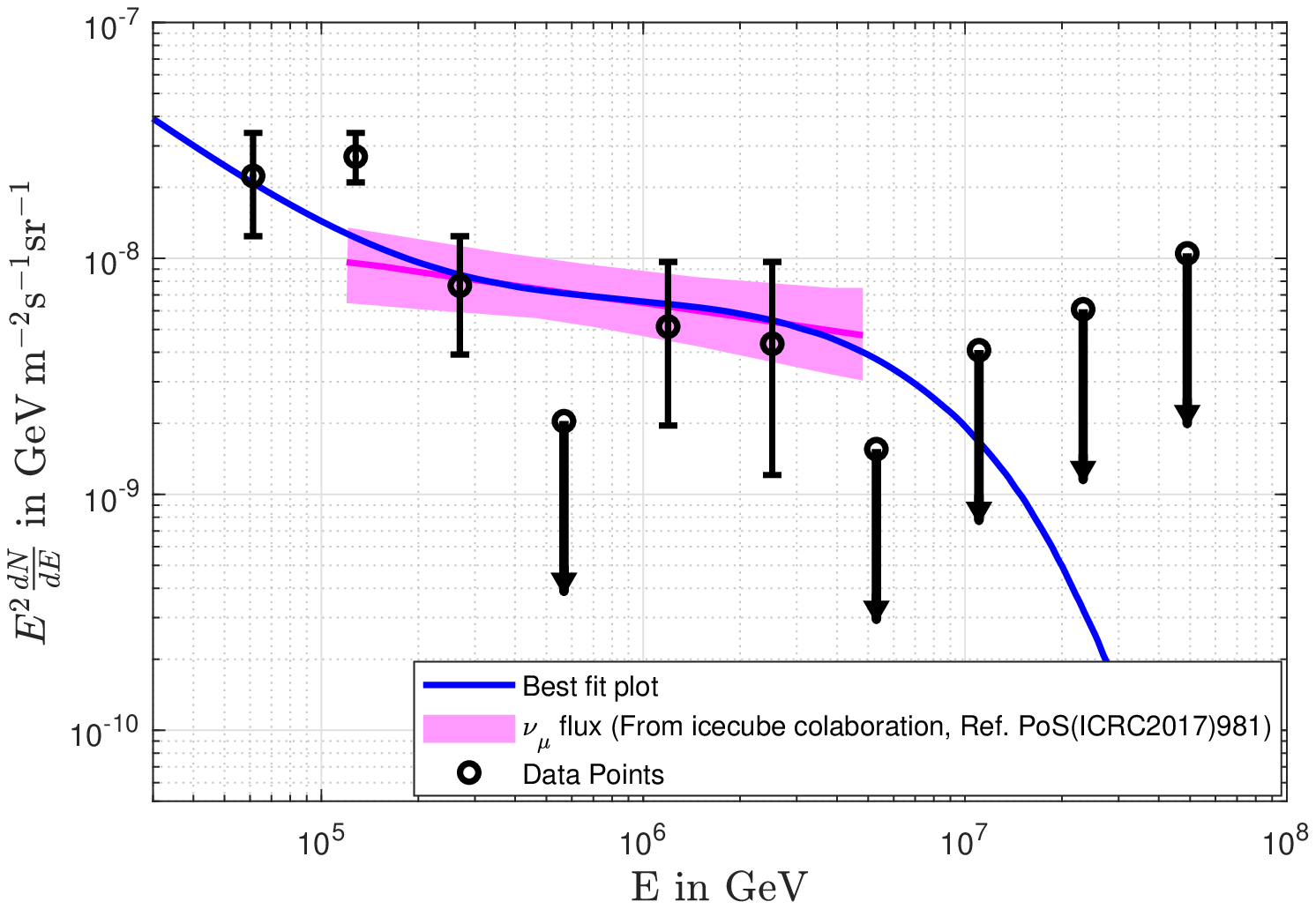}}
\subfigure []{
\includegraphics[height=6.0 cm, width=6.0 cm,angle=0]{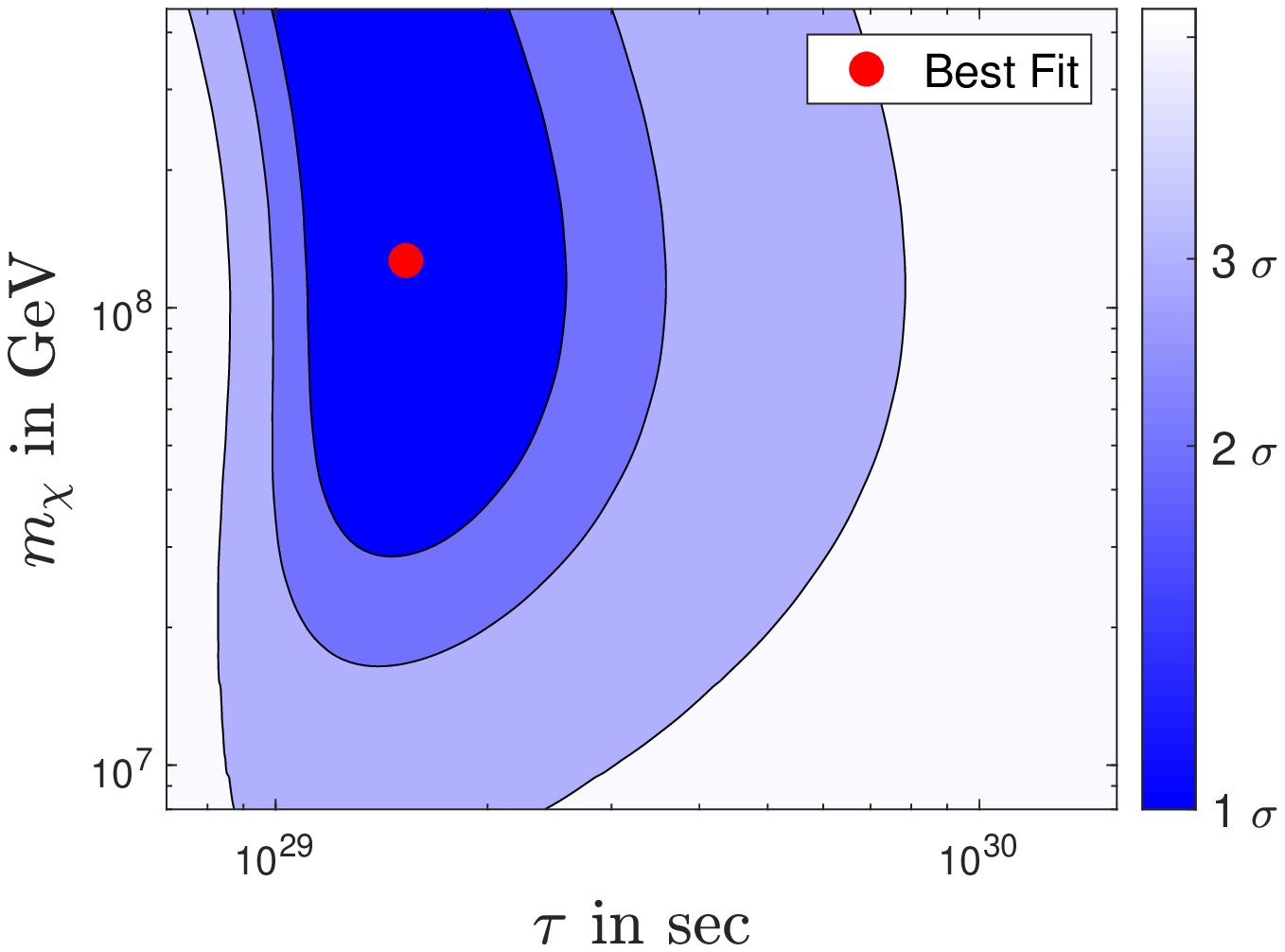}}
\caption{(a)flux, (b) contour by considering first 5 points (except the last
three points with large error bar), hadronic+astro channels
(Eq. (\ref{flux}) is considered as the astrophysical flux). 
See text for details.}
\label{figr6}
\end{figure}

%\begin{figure}[h!]
%\centering
%\subfigure[]{
%\includegraphics[height=6.0 cm, width=6.0 cm,angle=0]{noastropts_all(ic)_flux.eps}}
%\subfigure []{
%\includegraphics[height=6.0 cm, width=6.0 cm,angle=0]{noastropts_all(ic)_contour.eps}}
%\caption{(a)flux, (b) contour by considering no astro points (first two pints
%in ICRC paper are excluded), all channels
%(IceCube flux as the astrophysical flux). See text for details.}
%\label{figr2}
%\end{figure}

%\begin{figure}[h!]
%\centering
%\subfigure[]{
%\includegraphics[height=6.0 cm, width=6.0 cm,angle=0]{noastropts_all(ch)_flux.eps}}
%\subfigure []{
%\includegraphics[height=6.0 cm, width=6.0 cm,angle=0]{noastropts_all(ch)_contour.eps}}
%\caption{(a)flux, (b) contour by considering no astro points (first two pints
%in ICRC paper are excluded), all channels
%(chianese flux as the astrophysical flux). See text for details.}
%\label{figr2}
%\end{figure}

%\begin{figure}[h!]
%\centering
%\subfigure[]{
%\includegraphics[height=6.0 cm, width=6.0 cm,angle=0]{noastropts_leptohadro_flux.eps}}
%\subfigure[]{
%\includegraphics[height=6.0 cm, width=6.0 cm,angle=0]{noastropts_leptohadro_contour.eps}}
%\caption{(a)flux, (b) contour by considering no astro points (first two pints
%in ICRC paper are excluded), leptonic+hadroni channels . See text for details.}
%\label{figr2}
%\end{figure}

\begin{figure}[h!]
\centering
{\includegraphics[height=6.0 cm,angle=0]{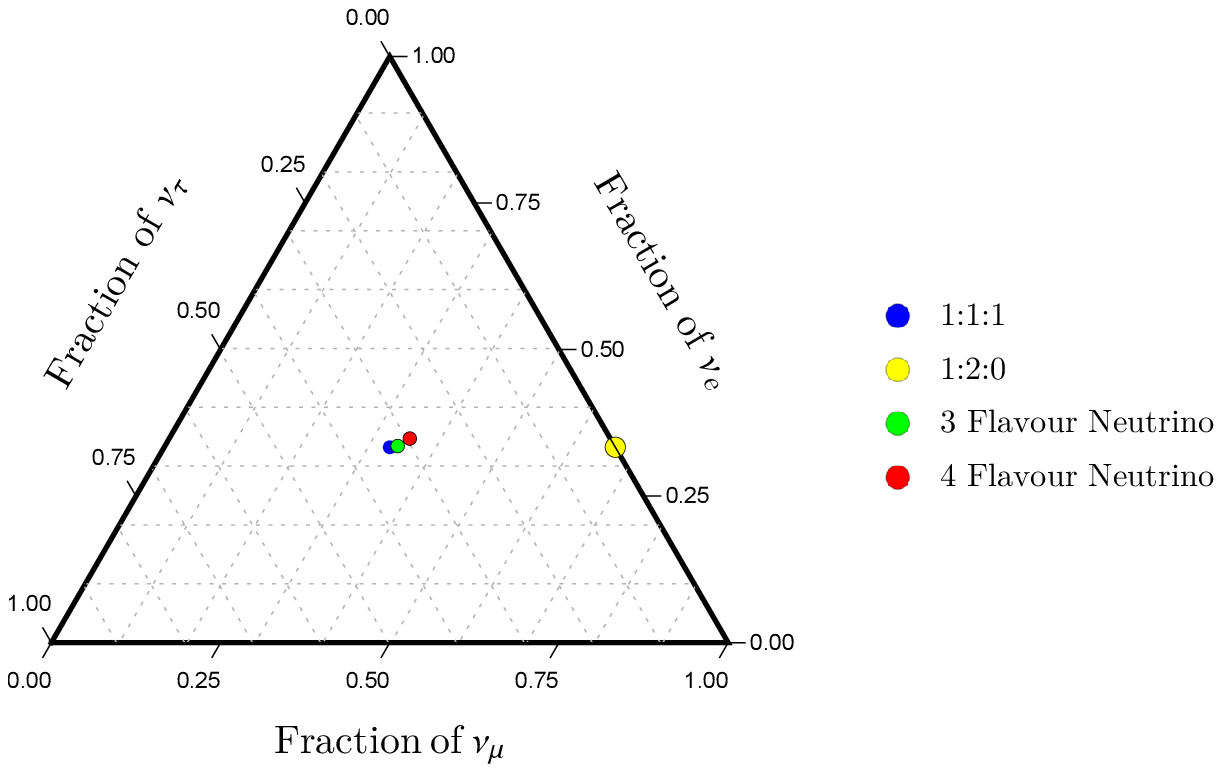}}
\caption{Ternary plot showing the neutrino flavour ratios on arriving 
the Earth from high energy sources.}
\label{fig2}
\end{figure}

%\begin{figure}[h!]
%\centering
%\subfigure[]{
%\includegraphics[height=6.0 cm, width=6.0 cm,angle=0]{flux_3_flavour.eps}}
%\subfigure []{
%\includegraphics[height=6.0 cm, width=6.0 cm,angle=0]{flux_4_flavour.eps}}
%\caption{See text for details.}
%\label{figr2}
%\end{figure}
  
%\begin{figure}[h!]
%\centering
%\subfigure[]{
%\includegraphics[height=6.0 cm, width=6.0 cm,angle=0]{contour_3_flavour.eps}}
%\subfigure []{
%\includegraphics[height=6.0 cm, width=6.0 cm,angle=0]{contour_4_flavour.eps}}
%\caption{See text for details.}
%\label{figr2}
%\end{figure}

%\begin{figure}[h!]
%\centering
%{\includegraphics[height=6.0 cm, width=7.5 cm,angle=0]{ternary_plot.eps}}
%{\includegraphics[height=6.0 cm,angle=0]{ternary_plot.eps}}
%\caption{See text for details.}
%\label{fig2}
%\end{figure}
\section{Discussions and Conclusions}
In this work we consider the UHE neutrino events reported by IC Collaboration 
in the energy range $\sim$ 60  TeV - $\sim$ 50 PeV. We propose that 
these UHE neutrinos originate from the decay of a supermassive dark matter 
that could have produced by the process of gravitational production in the 
early Universe. We make a $\chi^2$ analysis of the IC data in the energy 
range mentioned above by considering the neutrino flux from such dark matter 
decay as also from the possible astrophysical origin. For the computation 
of the astrophysical flux we consider a Waxman-Bahcall type power law as also 
the power law $\sim E^{-2.9}$ given by the IC Collaboration from their analysis.

From the present calculations, it appears that the energy range of UHE 
neutrinos considered here from IC data has three regions. 
The lower energy range between 
$\sim$ 60 TeV to $\sim$ 120 TeV represented by two event data points (Fig. 1) 
appears to be 
consistent with the flux of astrophysical origin, while for the higher energy 
range between $\sim 1.2 \times 10^5$ GeV - $\sim 5 \times 10^7$ GeV, the 
SHDM decay 
consideration of neutrino production pursued in this work, appear to describe 
well. Within this energy range again, the neutrino events in the range 
between $\sim 1.2 \times 10^5$ GeV - $\sim 5 \times 10^6$ GeV can be very 
well fitted 
with the neutrino flux following the hadronic channel of the dark matter 
decay. This hadronic channel however cannot describe at all the possible 
events in the higher energy region appearing between 
$\sim 5 \times 10^6$ GeV - $\sim 5 \times 10^7$ GeV. Although the event data 
is not very 
specific in this region with only the upper limit of four possible events 
are given by IC (Fig. 1) but the apparent nature of the flux appears 
to be very different than what is obtained for the range 
$\sim 1.2 \times 10^5$ GeV - $\sim 5 \times 10^7$ GeV (well described in this 
analysis by hadronic channel of SHDM decay). But we find that when the 
leptonic channel decay of SHDM is included in our analysis, 
the apparent nature of neutrino flux in this high energy regime can be
well represented.

Thus from our analyses it appears that the UHE neutrino signals in the enrgy 
range $\sim 1.2 \times 10^5$ GeV - $\sim 5 \times 10^7$ GeV reported by IceCube could have 
originated from the decay of superheavy dark matter.

{\bf Acknowledgments} : One of the authors (M.P.) thanks the DST-INSPIRE
fellowship (DST/INSPIRE/FELLOWSHIP/IF160004) grant by Department of Science and Technology (DST), Govt. of India. One of the authors (A.H.)
wishes to acknowledge the support received from St.Xavier’s College, kolkata
Central Research Facility and thanks the University Grant
Commission (UGC) of the Government of India, for providing financial support,
in the form of UGC-CSIR NET-JRF.

\end{document}